\documentclass[twocolumn,trackchanges]{aastex631}

\usepackage{amsmath}
\usepackage{relsize}
\usepackage[T1]{fontenc}
\usepackage{threeparttable}
\DeclareUnicodeCharacter{2212}{-}

\shorttitle{Exotic Stellar Populations in NGC\,2818}
\shortauthors{Rani et al.}
\graphicspath{{./}{figures/}}

\begin{document}

\title{UOCS-IX. AstroSat/UVIT study of the open cluster NGC\,2818: Blue Stragglers, Yellow Stragglers, Planetary Nebula, and their membership}


\correspondingauthor{Sharmila Rani}
\email{sharmila.rani@iiap.res.in}

\author[0000-0003-4233-3180]{Sharmila Rani}
\affiliation{Indian Institute of Astrophysics, Bangalore, 560034,  India}
\affiliation{Pondicherry University, R.V. Nagar, Kalapet, 605014, Puducherry, India}
\author[0000-0001-5812-1516]{Gajendra Pandey}
\affiliation{Indian Institute of Astrophysics, Bangalore, 560034,  India}
\author[0000-0003-4612-620X]{Annapurni Subramaniam}
\affiliation{Indian Institute of Astrophysics, Bangalore, 560034,  India}
\author[0000-0002-8414-8541]{N. Kameswara Rao}
\affiliation{Indian Institute of Astrophysics, Bangalore, 560034, India}





\begin{abstract}
We present the first far-UV (FUV) imaging results of the intermediate-age Galactic open cluster NGC\,2818 that has a Planetary nebula (PN) within the field using images taken from the Ultra-violet Imaging Telescope (UVIT) aboard \textit{AstroSat}. We identify cluster members by combining UVIT-detected sources with \textit{Gaia} EDR3 data. We detect four bright and hot blue straggler stars (BSSs) and two yellow straggler stars (YSSs) based on their location in the optical and FUV-optical color-magnitude diagrams. Based on the parameters estimated using Spectral Energy Distribution (SED), we infer that BSSs are either collisional products or might have undetectable white dwarf (WD) companions. Our photometric analysis of YSSs confirms their binarity, consistent with the spectroscopic results. We find YSSs to be formed through a mass-transfer scenario and the hot components are likely to be A-type subdwarfs. A comparison of the radial velocity (RV), Gaia EDR3 proper-motion of the PN with the cluster, and reddening towards the PN and the cluster does not rule out the membership of the PN. Comparing the central star’s position with theoretical pAGB models suggest that it has already entered the WD cooling phase, and its mass is deduced to be $\sim 0.66 M_{\odot}$. The corresponding progenitor mass turns out to be $\sim 2.1 M_{\odot}$, comparable to the turn-off mass of the cluster, implying that the progenitor could have formed in the cluster. We suggest that the NGC\,2818 might be one of the few known clusters to host a PN, providing a unique opportunity to test stellar evolution models.

\end{abstract}


\section{Introduction} \label{sec:intro}
Open clusters (OCs) are ideal laboratories to probe
the structure and history of the Galactic disk. They
are also test-beds to study the formation and evolution
of single and binary stellar populations. Dynamical interactions of stellar populations in star clusters lead to binaries and the formation of exotic stellar populations such as blue straggler stars (BSSs), yellow straggler stars (YSSs), and cataclysmic variables. These systems, as well as the end products of stellar evolution, such as hot white dwarfs (WDs), emit the bulk of their energy in the ultraviolet (UV) regime. UV observations of OCs are crucial to detect and understand the properties of the hot stellar populations, as highlighted in \cite{1997ApJ...481L..93L} and \cite{2008ApJ...683.1006K}.

One of the intriguing products of stellar interactions
in the OCs are BSSs whose origin and evolution are still
debated \citep{Boffin2015}. 
As these stars appear brighter and bluer than the stars located in the MS turn-off region of the cluster  color-magnitude diagram (CMD), they are expected to be more massive than the turn-off stars. To explain the mass gain and rejuvenation of these objects, the main formation scenarios proposed are, direct collisions or spiraling in of binary stars resulting in mergers \citep{1976ApL....17...87H}, or mass-transfer activity in close-binary systems \citep{1964MNRAS.128..147M}.  The dynamical evolution of hierarchical triple systems leading to the merger of an inner binary via the Kozai mechanism \citep{1999ASPC..169..432I, 2009ApJ...697.1048P} is another possible mechanism. Observational studies of BSSs suggest that a combination of all the formation channels are prevalent, and has a dependence on their environment, as they are found in a variety of stellar environments such as OCs \citep{2007A&A...463..789A, 2006A&A...459..489D}, globular clusters (GCs) \citep{2012Natur.492..393F}, the Galactic field \citep{2015ApJ...801..116S}, and dwarf galaxies \citep{2012ASPC..458..339S}. Thus, studying BSSs can provide information about the dynamical history of the cluster, the role of the dynamics on binary evolution, the frequency of binary systems, and the contribution of binaries to cluster evolution.
Member stars that are redder than the BSSs and brighter than the sub-giants found in the CMDs of OCs and GCs are considered as evolved BSSs, and are known as yellow straggler stars (YSSs) ( See \citealp{2018MNRAS.481..226S} and references therein). 

There are only a few OCs in our Galaxy known to harbor
Planetary nebulae (PNe). PNe are classically considered to represent the late stages in the stellar evolution of all the low as well as intermediate-mass stars with a mass range of 0.8$-$8 $M_{\odot}$ \citep{2000A&A...363..647W}. As the evolutionary lifetime of PNe are short (around $10^{3}-10^{5}$ years, depending on the mass of the progenitor) when compared to other evolutionary phases, especially when the number of evolved stars present in OCs are small, PNe as members of OCs are rare and are not expected in young OCs.  
Objects in this short-lived phase are critically important to our understanding of the physical processes and steps that transform stars into their remnants. They allow us to test the theory of stellar evolution, including the physics of nucleosynthesis and the relation between a star's initial mass and its white dwarf (WD) remnant \citep{2014RMxAA..50..203K}. Moreover, the chemical composition of the PNe can provide information about the dredge-up of chemical elements, which is expected to depend on the star's initial mass and composition. Finding a planetary nebula (PN) as a member of an OC gives us an excellent opportunity to better characterize and constrain its crucial parameters, such as distance, reddening, and age.

 NGC\,2818, has the unique distinction of being one of the two galactic OCs probably associated with a PN, and interestingly, the name NGC\,2818 is assigned to both an OC and a PN. Most importantly, the membership of the PN to the OC is still debated. In this study, we analyze both the cluster and the PN, NGC\,2818.


Here we present the results of the UV imaging of  NGC\,2818 (both PN and OC) in four far-UV (FUV) filters using the ultraviolet imaging telescope (UVIT) on {\it AstroSat}. Our main aims are: (1) to identify and characterize the blue and yellow straggler stars in the cluster to shed light on their formation and evolution and (2) to characterize the central star of the PN (CSPN) to investigate its association with the cluster. The age of this cluster is estimated to be $\sim$800 Myr, and the reddening of the cluster is E(B$-$V) = 0.2\,mag \citep{2021MNRAS.502.4350S}. This cluster is located at a distance of $3250\pm300$ pc and the metallicity is found to be solar \citep{2021MNRAS.502.4350S}.\\

NGC\,2818 is one of the OCs that shows an extended main-sequence turn-off (eMSTO) phenomenon \citep{2018MNRAS.480.3739B}, where the cluster MS is extended in the CMD more than what is expected from a simple stellar population with conventional evolutionary history. It has been demonstrated that stellar rotation is the most probable cause of this phenomenon \citep{2009MNRAS.398L..11B, 2015ApJ...807...24B, 2015MNRAS.453.2070N, 2016MNRAS.457..809C, 2019ApJ...887..199G}. A spectroscopic study by \cite{2018MNRAS.480.3739B} showed that, in NGC\,2818, stellar rotation is indeed linked to the stars' position on the MSTO of the CMD made using the Gaia magnitudes (G) and color (Gbp$-$Grp), such that rapidly rotating stars preferentially lie on the red side of the eMSTO.
However, the color range (Gbp$-$Grp) in optical CMD is relatively small, whereas a larger color range is seen in UV colors, and it is expected that the rotational effects are more prominently displayed in UV colors mainly because of their sensitivity to surface (effective) temperature changes. 
This study also explores the correlation between the  colors derived from UVIT FUV filters and stellar rotation.

The layout of this paper is as follows. In section~\ref{sec:data}, we describe the observations, data reduction, and analysis methods. 
In Section~\ref{sec:members}, we present proper-motion-based membership information using \textit{Gaia} EDR3 data for cluster stars and PN. Section~\ref{sec:CMDs} presents the selection of BSSs and YSSs from the observed UV and Optical CMDs, including the stellar rotation effects on CMDs. In Sections~\ref{sec:SEDs} and ~\ref{sec:status}, we describe the properties of BSSs, and YSSs derived from the UVIT photometry along with GALEX, \textit{Gaia} and ground-based photometry and their evolutionary status. 
A detailed discussion of all results is provided in Section~\ref{sec:dis}. Finally, in Section~\ref{sec:summary}, we summarize our main results and conclusions.

\section{Observational Data and Analysis} 
\label{sec:data}
\subsection{UVIT Data}
  \label{subsec:uvitdata}
  In order to probe the nature of the exotic stellar populations in NGC\,2818, we use data acquired with the UVIT instrument on board the Indian multiwavelength astronomy satellite \textit{AstroSat}. UVIT produces images of the sky in far-UV (FUV), near-UV (NUV), and visible, simultaneously, over a circular field-of-view of 28$\arcmin$ diameter with a spatial resolution of $\sim 1\farcs5$ in both FUV and NUV channels. More details about the telescope, its initial and new calibration, and its results are described in detail by \cite{2017JApA...38...28T, 2020AJ....159..158T}. The derived magnitudes of the stellar sources observed with the UVIT filters are in the AB magnitude system.\\
  
\begin{figure*}[!htb]
    \centering
	\includegraphics[scale=0.55]{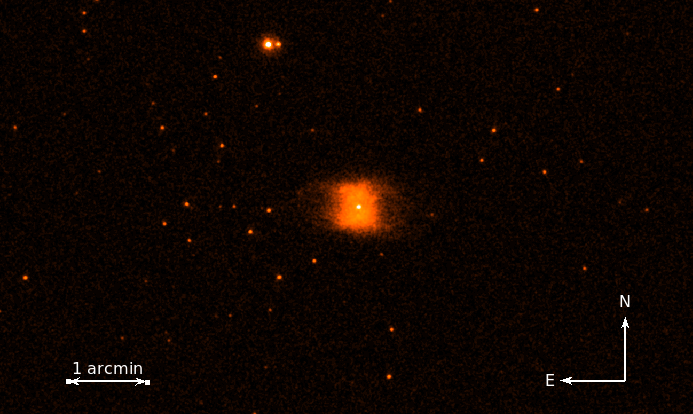}
    \caption{UVIT color image of OC NGC\,2818 in FUV F148W channel. Here orange color depicts the FUV detections. The extended structure in this image represents the PN NGC\,2818. North is up, and east is left in the image.}
    \label{fuvimage}
    \end{figure*}

  
The observations of NGC\,2818 used in this work were made in two epochs, first on 21st December 2018 (Prop: A05$\_$196 $-$P.I: N. K. Rao), and the second on 11th June 2020 (Prop: A09$\_$047 $-$P.I: N. K. Rao). In the first epoch, the observations were carried out in three FUV filters (F154W, F169M, and F172M), and in the second, observations were performed with deep exposures in four FUV filters (F148W, F154W, F169M, and F172M). The observations are carried out in several orbits in order to complete the allotted exposure times in given filters. We utilize a customized software package, CCDLAB (\citealp{2017PASP..129k5002P}), to correct for the geometric distortion, flat field, spacecraft drift and create images for each orbit. Then, the orbit-wise images were co-aligned and combined to generate science-ready images in order to get a better signal-to-noise ratio. Further analysis was done using these final science-ready images to obtain the magnitudes of the sources detected with UVIT. The details of the UVIT observations of NGC\,2818 used in this analysis are tabulated in Table~\ref{tab1}. In Figure~\ref{fuvimage}, we show the UVIT image of the cluster taken in the FUV F148W band where the orange color depicts FUV detections. This image exhibits an extended structure displaying the beautiful PN NGC\,2818, where the central star can be seen in the FUV. 

	 \begin{figure*}
	     \centering
	     \includegraphics[scale=0.225]{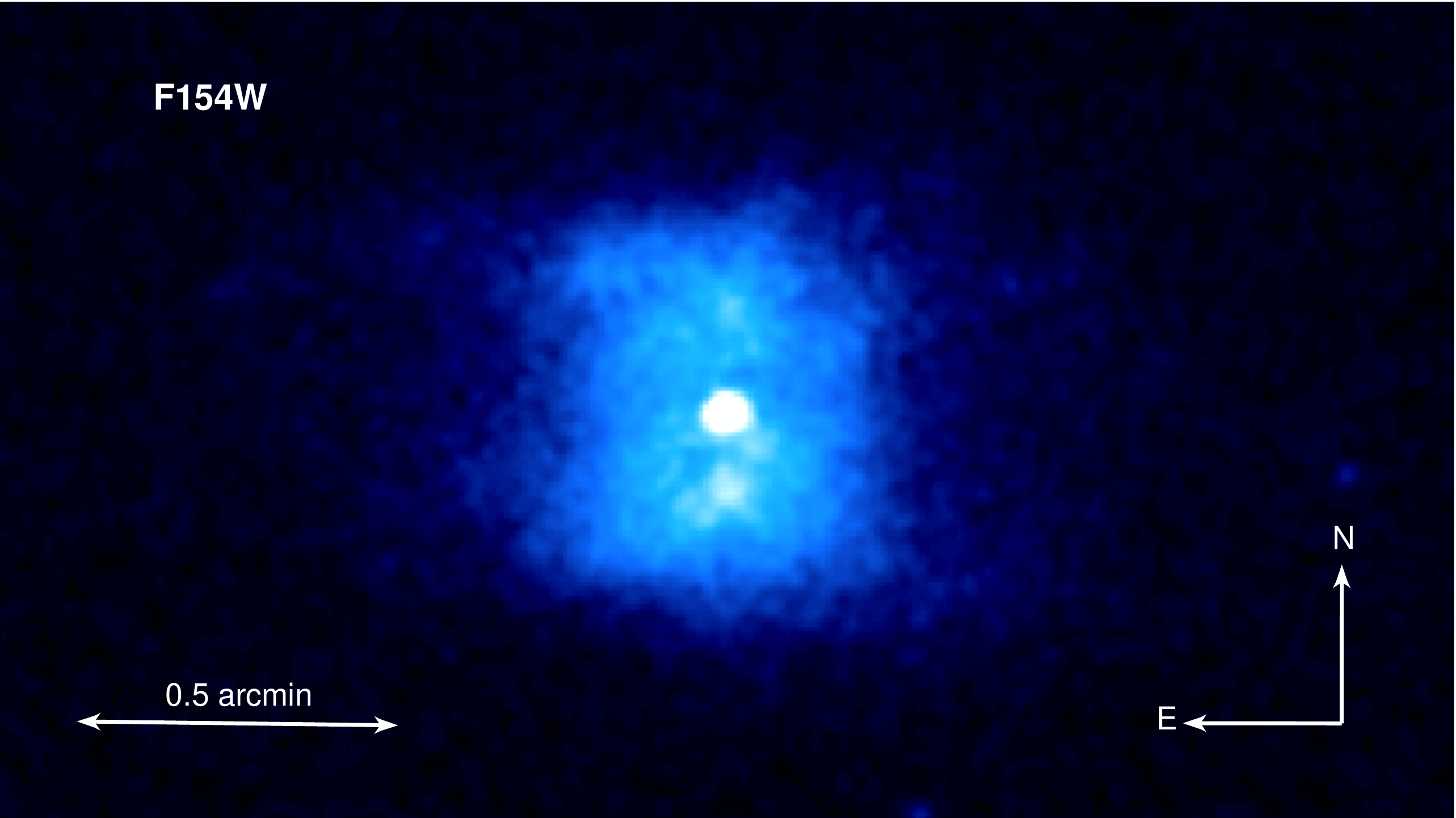}
	     \includegraphics[scale=0.225]{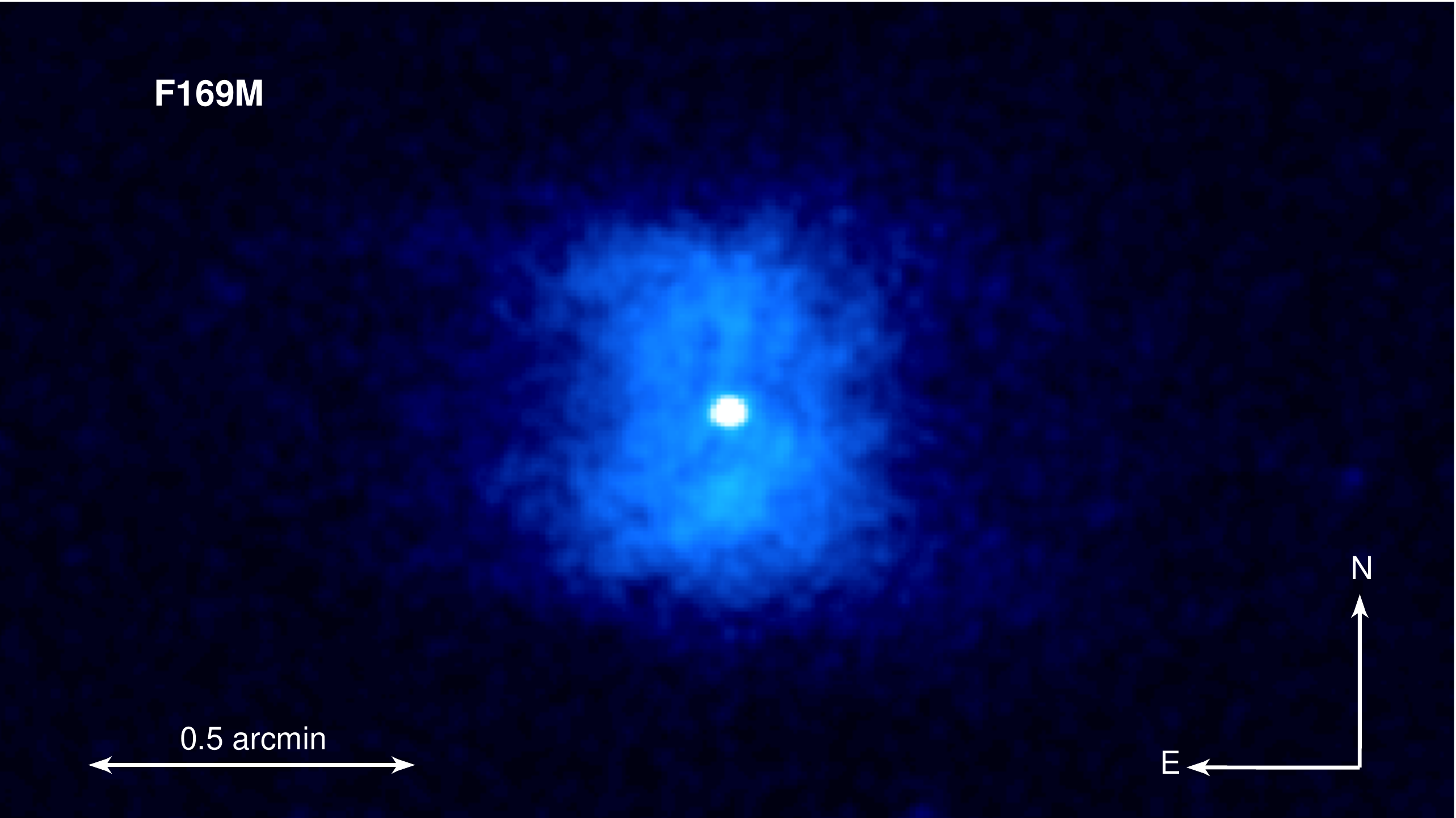}
	     \includegraphics[scale=0.225]{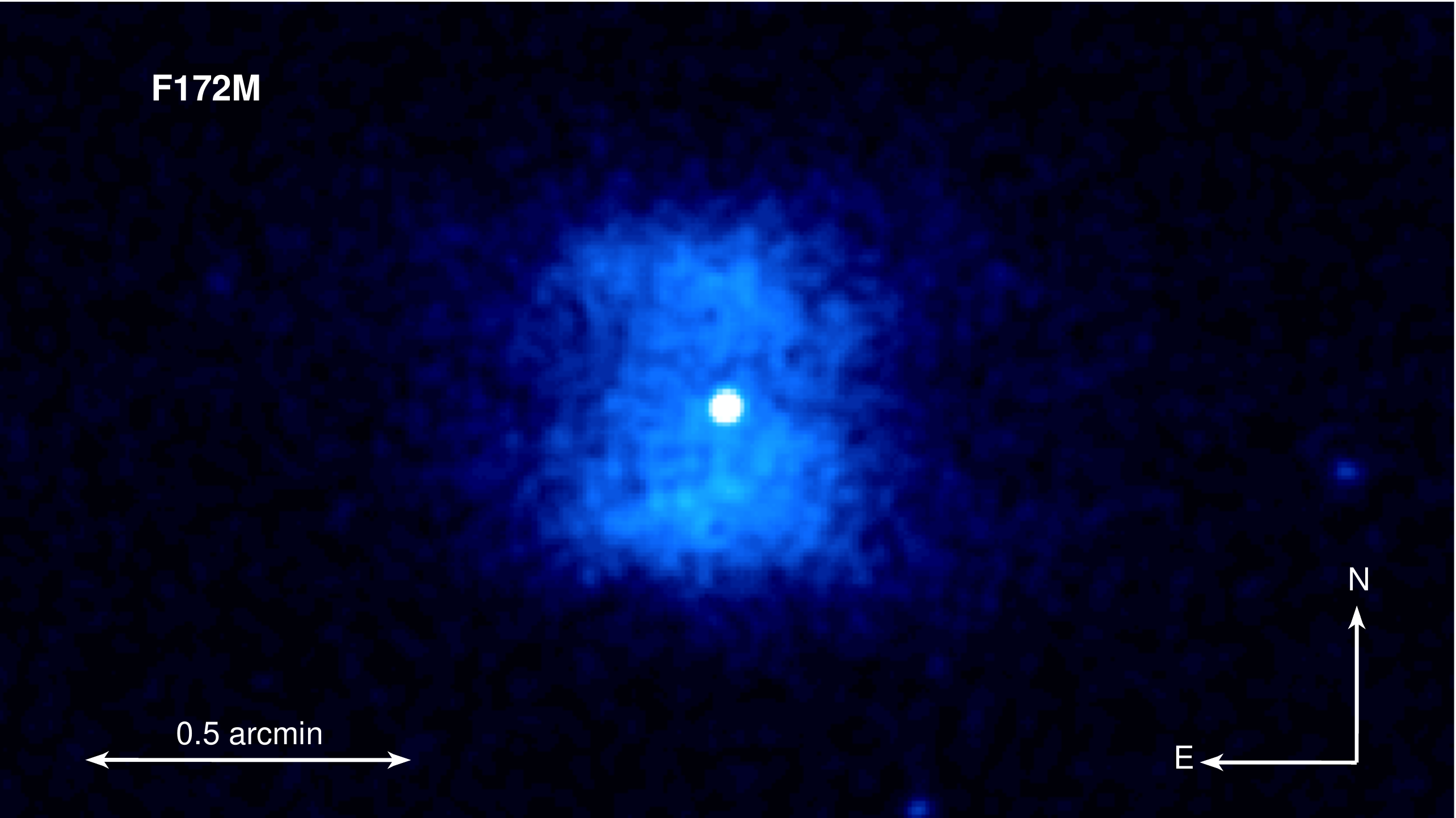}
	     \caption{UVIT/FUV images of PN NGC\,2818 in three filters: F154W, F169M, and F172M.}
	     \label{fig:my_label}
	 \end{figure*}
	 \begin{table*}[!htb]
    \centering
    \begin{threeparttable}[1pt]
	\caption{List of the FUV observations of NGC\,2818 obtained with UVIT in two epochs used in this work. 
	The last column lists the extinction value computed in each FUV filter using \cite{1999PASP..111...63F} law of extinction.}
	\label{tab1}
	\begin{tabular}{lcccccc} 
		\hline
		\hline
		 Filter & $\lambda_{mean}$ & $\Delta\lambda$  & ZP  & \multicolumn{2}{c}{t$_{\rm exp}$ (sec)} &  A$_{\lambda}$\\
		  & ({\AA}) & ({\AA}) & (AB mag) & (1st epoch)  &  (2nd epoch) & (mag)\\
		\hline
		 F148W & 1481 & 500 & 18.09 & - & 1736 & 1.58\\
		 F154W & 1541 & 380 & 17.77 & 1491 & 2877 & 1.55\\
		 F169M & 1608 & 290 & 17.41 & 1715 & 1999 & 1.54 \\
		 F172M & 1717 & 125 & 16.27 & 1903 & 2878 & 1.51\\
		 \hline
	\end{tabular}
	\end{threeparttable}
  \end{table*}
  
  \subsection{Photometry}
  \label{sec:phot}
\begin{figure}[!htb]
  \hspace*{-0.6cm} 
	\includegraphics[height=8.5cm,width=9cm]{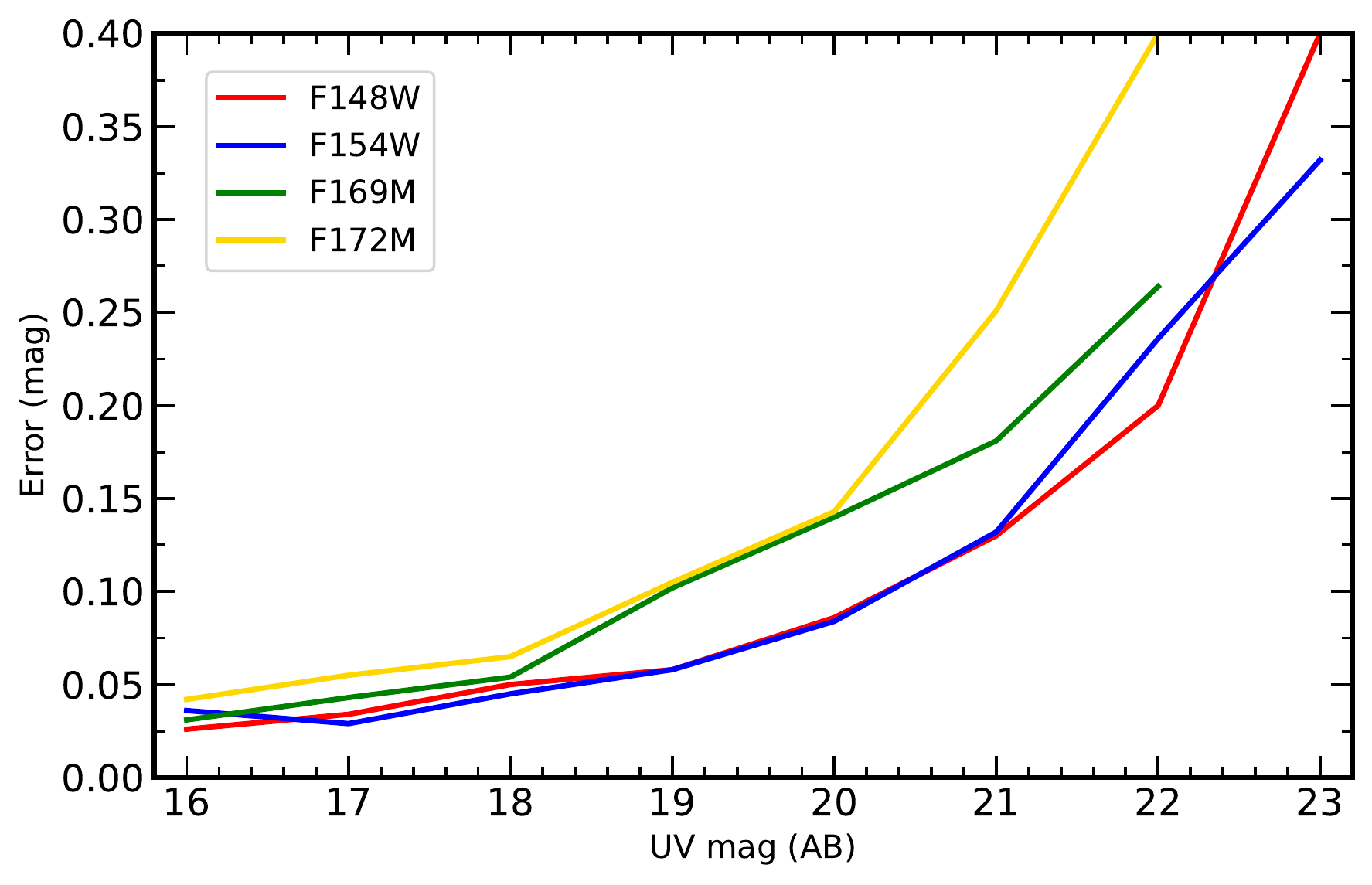}
     \caption{PSF-fit errors (median) as a function of magnitude for our UVIT observations of NGC\,2818 in all FUV bandpasses. 
     }
    \label{psferr}
\end{figure}

To extract the magnitudes of detected stars in all FUV images, we have carried out the point spread function (PSF) photometry using the IRAF/NOAO package DAOPHOT \citep{1987PASP...99..191S}. The steps taken to obtain the magnitude of the sources are as follows: First, the stars are located in the image using the DAOFIND task in IRAF. Further, we used the PHOT task to perform the aperture photometry. To construct the model PSF using the PSF task, bright and isolated stars are selected in the image using the PSTSELECT task. The average PSF of the stars in all FUV images is $\sim 1\farcs2$. The ALLSTAR task is used to fit the model PSF to all the detected stars in the image to obtain the PSF-fitted magnitudes. The PSF magnitudes were converted to aperture photometry scale using the PSF correction further followed by aperture correction, estimated using the curve of growth analysis by choosing isolated bright stars in the field. Finally, the saturation correction, in order to account for more than one photon per frame, was applied to the obtained magnitudes in UVIT filters. All steps to perform the saturation correction are described in detail in \cite{2017JApA...38...28T}. The extracted instrumental magnitudes are calibrated into the AB magnitude system using the zero points (ZP) reported in the recently published calibration paper \citep{2020AJ....159..158T}. Figure~\ref{psferr} shows the PSF-fit error (median) as a function of magnitude in four FUV filters for profound observations. We have detected stars up to $\sim$ 22\,mag with PSF-fit errors less than 0.3\,mag in all FUV filters and considered them for further analysis in the paper.\\
To apply the extinction and reddening correction to the derived UVIT magnitudes of all detected stars, we adopted the reddening, E(B$-$V) = 0.2\,mag mentioned in the \cite{2021MNRAS.502.4350S}. The ratio of total-to-selective extinction, $R_{V}$ = 3.1 for the Milky Way, was taken from \cite{1958AJ.....63..201W} to calculate the extinction value in the visual band ($A_V$). We used the Fitzpatrick extinction law \citep{1999PASP..111...63F} to compute extinction coefficients $A_\lambda$ for all UVIT filters, as listed in Table~\ref{tab1}.

\subsection{Other Catalogs}
\label{sec:archival}
This cluster was previously observed in UV, optical, and Infrared (IR) all-sky surveys with GALEX \citep{galexcatalog}, SDSS \citep{sdsscatalog}, APASS \citep{apasscatalog}, 2MASS \citep{2masscatalog}, and WISE \citep{allwisecatalog}, respectively. In this work, we combined the UVIT data with the multi-wavelength photometric catalog spanning a wavelength range from UV-IR.
We used the virtual observatory tool in VOSA to cross-match the UVIT-detected sources with the above-mentioned photometric catalogs \citep{2008A&A...492..277B}.

\begin{figure*}[!htb]
 \centering
	\includegraphics[width=0.72\columnwidth]{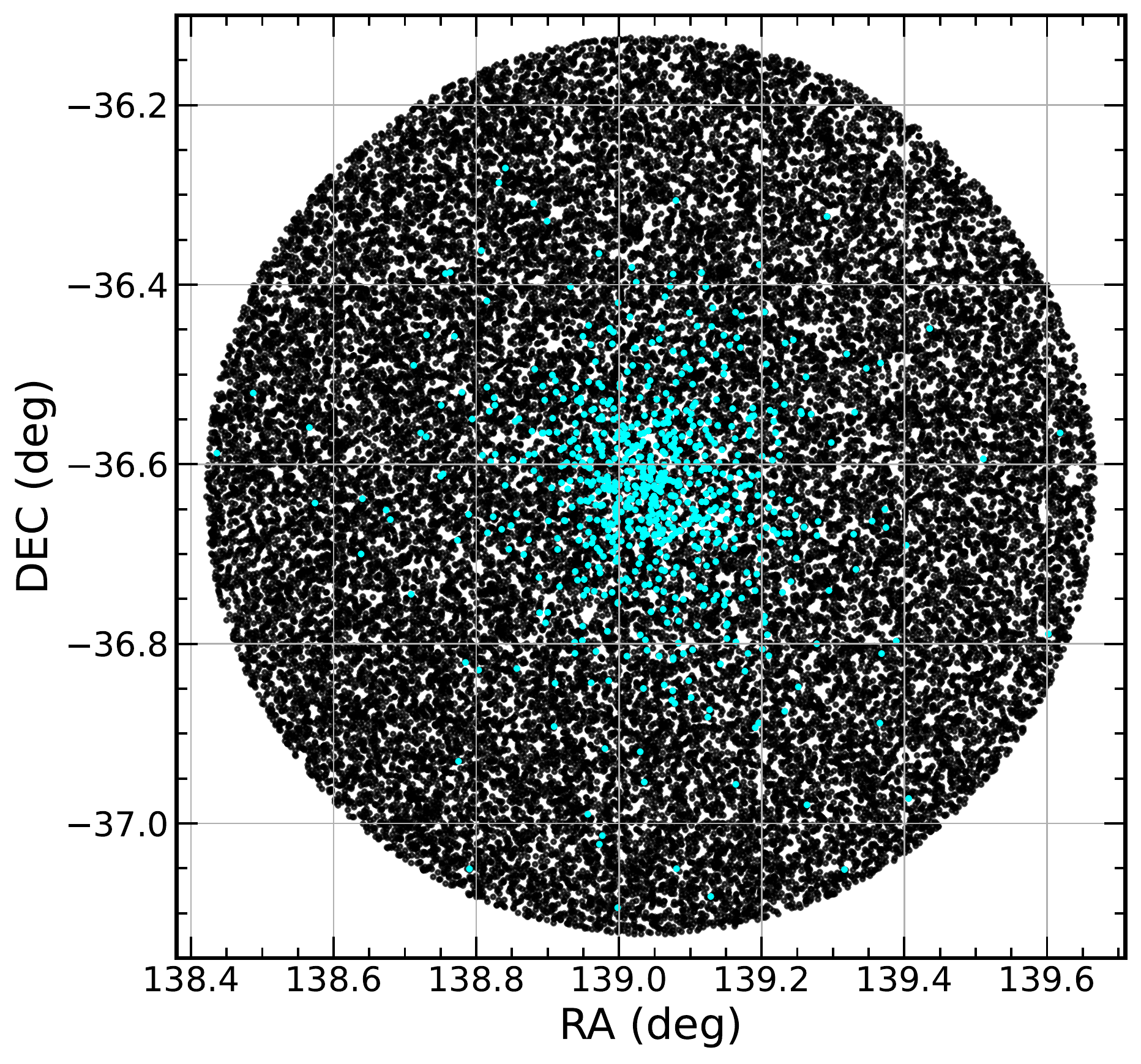}
	\includegraphics[width=0.68\columnwidth]{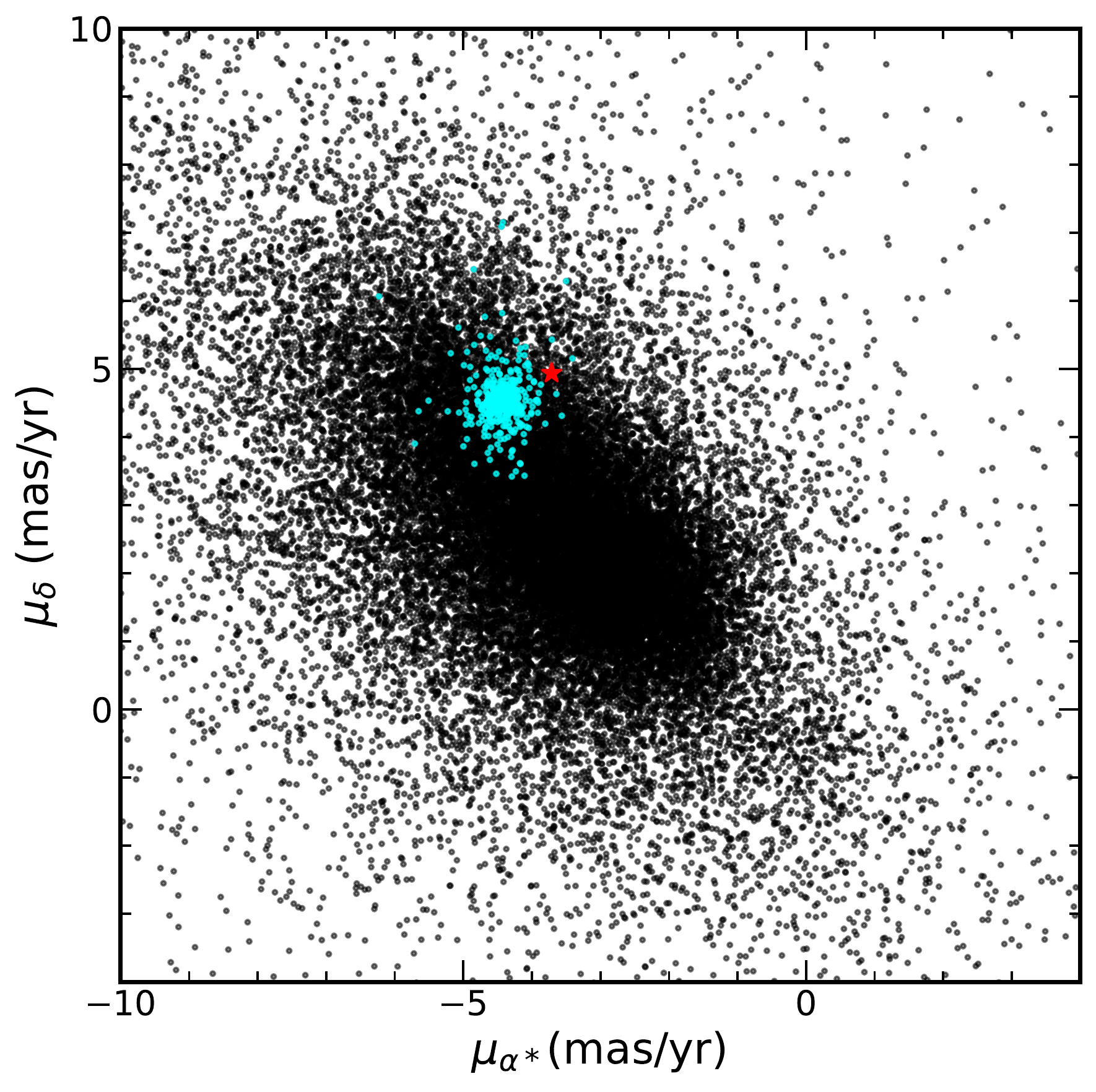}
	\includegraphics[width=0.685\columnwidth]{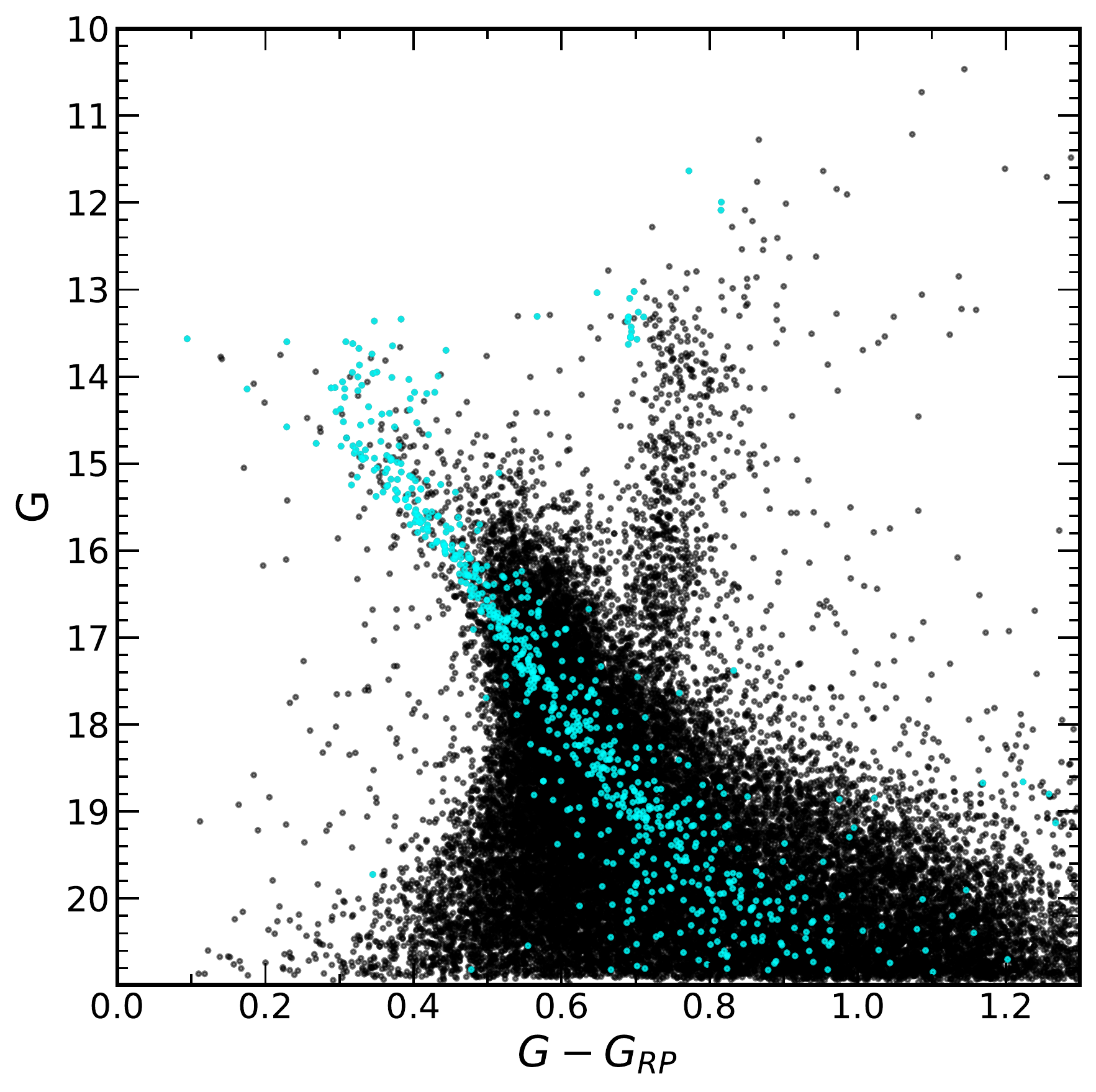}
     \caption{In three panels from left to right, PM members of the cluster are shown with cyan dots, and the remaining \textit{Gaia} EDR3 sample marked with black dots represents field stars. Left Panel: position in the sky; Middle Panel: Vector Point Diagram (VPD); Right Panel: \textit{Gaia} Optical CMD.}
    \label{gaiapm}
\end{figure*}

\section{Membership Determination} 
\label{sec:members}
We employed the \textit{Gaia} early data release 3 (EDR3) catalog that provides data with unprecedented precision
to identify the cluster members. In particular, it provides the complete 5-parameter astrometric solution (positions, proper motions, and parallaxes) and magnitudes in its three photometric bands (G, G$_{BP}$, and G$_{RP}$) with a limiting magnitude of about G$\sim$21\,mag. 
To assign the proper motion (PM) membership probability (P$_{\mu}$) of all stars observed in the cluster, we first downloaded all detections located within a 30$\arcmin$ radius from the cluster's center. To include all possible members of the cluster, we opted to use a radius bigger than that provided by \cite{2013A&A...558A..53K} catalog. Then, we applied the data quality criteria to select the sources with a good astrometric solution. Stars are selected as follows: (i) we removed those with parallaxes that deviate by more than 3$\sigma$ from the expected parallax calculated using the previously known distance to the cluster, where $\sigma$ is the error in parallax given in \textit{Gaia} EDR3 catalog, (ii) we also removed the sources with renormalized unit weight error (RUWE) exceeding 1.2 as larger values of this parameter might lead to an unreliable astrometric solution \citep{2018A&A...616A...2L, 2021A&A...649A...3R}.\\

We made use of a probabilistic Gaussian mixture model (GMM) method to select cluster members and infer the intrinsic parameters of the distributions of both member and non-member stars. In this method, the distribution of sources in the vector-point diagram ($\mu_{\alpha}, \mu_{\delta}$) is modeled as a mixture of two Gaussian distributions, one for the cluster members and another one for the field sources. The details of this method are well described in \cite{2019MNRAS.484.2832V}.
The Gaussian probability distribution corresponding to the sum of two distributions is

\begin{equation}
\hspace{0cm}
    f(\mu|\overline{\mu_{i}}, \mathsmaller{\sum_{i}}) = \sum_{i=1}^{2} w_{i}\frac{exp\big[-1/2(\mu - \overline{\mu_{i}})^{T}\sum_{i}^{-1}(\mu - \overline{\mu_{i}})\big]}{2\pi\sqrt{det\sum_{i}}}
\end{equation}

\begin{equation}
\hspace{0cm}
    w_{i}\geq0, \indent \sum_{i=1}^{2} w_{i}= 1
\end{equation}
where $\mu$ is individual PM vector; $\overline{\mu_{i}}$ are field and cluster mean PMs; $\scriptstyle\sum$ is the symmetric covariance matrix; and $w_{i}$ are weights for the two Gaussian distributions. Full details of this method for the n-dimensional case are described in \citep{2019MNRAS.484.2832V}.\\

The initial guess for cluster PM $\mu_{\alpha}$ and $\mu_{\delta}$ values and internal velocity dispersion are taken from \citep{2020A&A...640A...1C}. We utilized GaiaTools\footnote{\url{https://github.com/GalacticDynamics-Oxford/GaiaTools}} to maximize the total log-likelihood of GMM and measure the mean PM and standard deviation of both the Gaussian distributions. The membership probabilities (MPs) of all the selected stars are calculated using the same technique simultaneously. The equations used to maximize the log-likelihood of GMM and estimate the MP of the $i^{th}$ star belonging to the $k^{th}$ component are given in appendix A in \cite{2019MNRAS.484.2832V}.

 The PM mean and standard deviations of the cluster distribution are computed to be $\mu_{\alpha}$ = -4.417 mas/yr and $\mu_{\delta}$ = 4.540 mas/yr, with $\sigma_{c}$ = 0.045 mas/yr. In Figure~\ref{gaiapm}, we show the position of stars in the sky, in the PM space known as vector point diagram (VPD), and in an optical CMD created using \textit{Gaia} filters. Cyan dots in all the plots depict the member stars belonging to the cluster, and black dots represent the field stars. 718 stars are identified as most likely cluster members with P$_{\mu}$>50$\%$ and considered for subsequent analysis. This method works well for a  distinguishable distribution of PM  for the field and cluster stars in the VPD. But, in this case, the PM of cluster stars are located well within the PM distribution of the field stars, suggesting a non-trivial identification of cluster members from field stars. Therefore, it is possible that stars with a lower membership probability than the above-mentioned limit might also be members of the cluster.
 
\subsection{Is PN a member of the cluster?}
\label{PNememb}
The membership of the PN with OC has been debated in several studies in the past. \cite{1972MNRAS.158...47T} found that PN NGC\,2818 is a member of the OC of the same name. \cite{1984ApJ...287..341D} presented the results of photometric as well as spectroscopic observations of the nebula to analyze its physical properties and chemical composition. He suggested that the nebula is probably associated with the star cluster. \cite{1989AJ.....98.2146P} analyzed this cluster using CCD UBV photometric data and assumed a physical association of the nebula with the cluster. \cite{1990JApA...11..151S} also suggested the association of the PN with the cluster from their CCD photometry of the cluster. However, \cite{2001A&A...375...30M} derived accurate heliocentric radial velocities for 12 cluster red giants to obtain a mean heliocentric radial velocity of Vhel = +20.7 $\pm$ 0.3 kms$^{-1}$, significantly different from the PN velocity of $−$1 $\pm$ 3 kms$^{−1}$ \citep{1988ApJ...334..862M}, suggesting that they are unrelated. Recently, \citep{2012ApJ...751..116V} reanalyzed the complex kinematics and morphology of the nebula using high-resolution Hubble Space Telescope (HST) archive imaging and high-dispersion spectroscopic data and determined a systemic heliocentric velocity of PN to be +26$\pm$2 kms$^{−1}$ in closer agreement with the OC, suggesting its membership. Moreover, based on its RV, H$\alpha$ surface brightness, and radius, \cite{2016MNRAS.455.1459F} concluded that the PN might be a cluster member. 

The \textit{Gaia} EDR3 trigonometric parallax for the central star of the nebula (CSPN) is  0.0319$\pm$0.21 mas, but it can be noted that the uncertainty in it is more than its value. So, it can not be used to obtain the distance to the nebula. The best estimate of the statistical distance is given by \citep{2016MNRAS.455.1459F} as 
3000$\pm$800 pc not too far from cluster distance of 3250$\pm$300 pc estimated by \cite{2021MNRAS.502.4350S}. \citep{2020A&A...640A...1C, 2020A&A...633A..99C} obtained the members of the several OCs, including NGC\,2818, using \textit{Gaia} DR2 PM data, and suggested that it is a non-member of the cluster. 


In our membership analysis, we have obtained the membership of the CSPN using the \textit{Gaia} EDR3 PM data. The PM in RA and DEC of the CSPN as listed in \textit{Gaia} EDR3 catalog is $\mu_{\alpha}$ = $-3.712 \pm 0.185$ mas/yr and $\mu_{\delta}$ = $4.94 \pm 0.18$ mas/yr. Its P$_{\mu}$ is estimated to be $\sim$11$\%$, indicating non-membership. Nevertheless, it can be noted from the location of the CSPN shown with the red star symbol in the VPD that it is lying close to the PM distribution of the cluster members (Cyan dots), implying that it is quite likely a member of the cluster. Statistically, it is lying within 3$\sigma$ of the mean PM of the cluster. We expect that the future \textit{Gaia} data release (\textit{Gaia} DR4) might give more precise and accurate PM measurements that can re-confirm its association with the cluster. Further, assuming both cluster and nebula at the same distance, we computed their true velocity using their already available RV and PM information. We found that the true velocity of the cluster and nebula turn out to be approximately the same (V$_{C}$ = $99.7 kms^{-1}$ \& V$_{PN}$ = $98.7 kms^{-1}$), implying that the values of the space velocity are similar. 
  
\subsubsection{Reddening towards the PN}
\label{sec:reddening}
Several estimates of extinction/reddening towards the cluster have been made since the initial investigation by \cite{1972MNRAS.158...47T} of E(B$-$V) of 0.22 mag, reconfirmed by \cite{1990JApA...11..151S} and recently refined by \cite{2021MNRAS.502.4350S}, to 0.20 mag. However, there are a few independent estimates of extinction towards the PN NGC\,2818. \cite{1984ApJ...287..341D} estimated it from the Balmer lines $H\alpha/H\beta$ ratio as 0.24$\pm$0.02 mag. \cite{1988A&A...197..266G} list a value of 0.20 mag, and \cite{2016MNRAS.455.1459F} estimated a value of 0.17$\pm$0.08 mag. We presently estimate E(B$-$V) value using free-free continuum flux and the nebular $H\beta$ flux. The flux density, $S_{\nu}$ at 5 GHz of the entire nebula, is measured by \cite{1995ApJS...98..659Z} as 33 mJy. The total $H\beta$ flux is estimated by \cite{1988A&A...197..266G} as $log F(H\beta)$ as -11.40 ($erg cm^{-2} s^{-1}$). Following \cite{1984ASSL..107.....P}, the expected ratio of $S_{\nu}$ to F($H\beta$) is given as

\begin{equation*}
    \frac{S(_{\nu})}{F(H\beta)} = 2.51\times10^{7}\times T_{e}^{0.53}\times(\nu)^{-0.1}\times Y Jy/ergcm^{-2}s^{-1}
\end{equation*}
 where $T_{e}$ is the electron temperature; $\nu$ is frequency in GHz; Y = (1 + $\frac{n(He^{+})}{n(H^{+})}$). The value of $\frac{n(He^{+})}{n(H^{+})}$ is $\sim 0.13$ assuming all He is in $He^{+}$ form. \cite{1984ApJ...287..341D} derived the $T_{e}[OIII]$ of 14,500$\pm$500 K. From the above relation, the $log F(H\beta)$ expected from the radio continuum is -11.07.  The equation from \cite{1975A&A....38..183M} used to compute the reddening is following: 
 
\begin{equation*}
    E(B-V) = \frac{1}{1.46}log\frac{F(H\beta)_{exp}}{F(H\beta)_{obs}}
\end{equation*}

Inserting the expected and observed $log F(H\beta)$ values in the above equation, we obtain the value of E(B$-$V) $\sim$0.23 mag. Thus, the extinction/reddening towards this cluster and nebula is of similar value.

From the comparison of distance, RV, PM, and extinction/reddening values of the cluster and nebula, we suggest a physical association of the PN with the OC.

\section{Color Magnitude Diagrams}
\label{sec:CMDs}

\begin{figure}[!htb]
\hspace{-0.7cm}
\includegraphics[width=1.1\columnwidth]{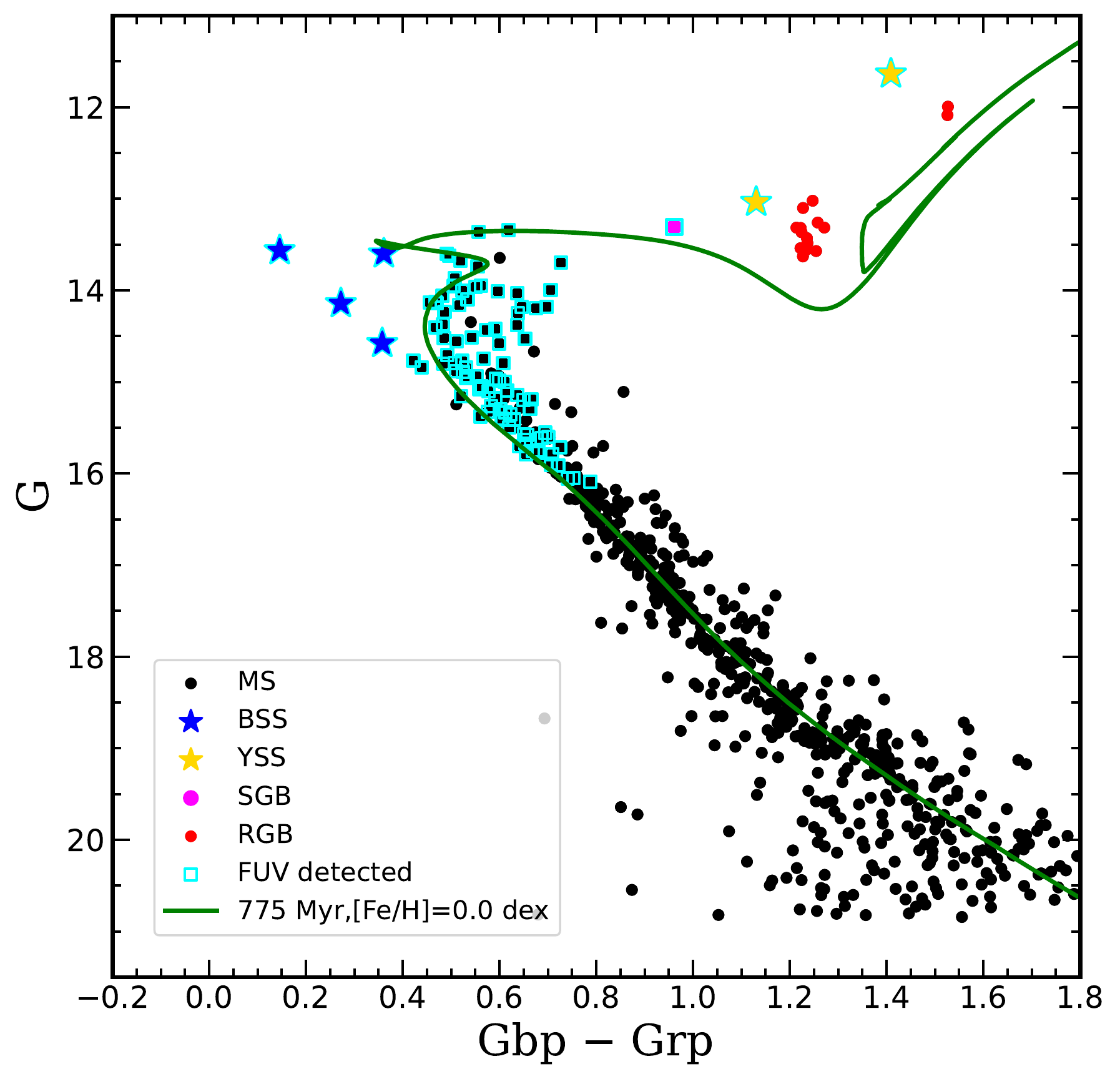}
 \caption{Optical CMD of the NGC\,2818, created using \textit{Gaia} EDR3 photometry. All filled symbols denote the stars with P$_{\mu}\geq50\%$. Blue-filled stars and yellow-filled stars are the selected blue and yellow straggler stars used for further cross-match with UVIT data, respectively. The stars detected in all FUV images are outlined with cyan-colored square and star symbols.
 The over-plotted green solid line represents the non-rotating MIST isochrone of solar metallicity and an age of 775 Myr, set at reddening, E(B$-$V)=0.2\,mag and distance modulus, (m$-$M)$_{V}$ = 12.56\,mag.}
 \label{optcmd}
\end{figure}

\subsection{Classification of Exotic sources}
\label{sec:class}
This section describes the classification and identification of exotic sources, such as BSSs and YSSs, expected to emit in the FUV. As mentioned in Section~\ref{sec:members}, we considered the probable cluster members with P$_{\mu}$>50$\%$ and created the PM-cleaned optical CMD (G$_{bp}$ - G$_{rp}$ vs. G) using the \textit{Gaia} filters shown in Figure~\ref{optcmd}. In this CMD, stars outlined with cyan color depict the various identified star populations in FUV images. \cite{2021A&A...650A..67R} presented a new proper-motion-cleaned catalog of BSSs in galactic OCs using \textit{Gaia} DR2 data. We cross-matched the \textit{Gaia} EDR3 cluster members with the BSS catalog to classify this population in the cluster. Out of five identified BSSs in NGC\,2818 by \cite{2021A&A...650A..67R}, we detected four BSSs. The remaining one BSS, not detected by us, is found to be a non-member of the cluster in our membership catalog and also falls outside the FoV of NGC\,2818 observed with UVIT in two epochs. \cite{2021MNRAS.507.1699J} also produced a catalog of BSSs in OCs using \textit{Gaia} DR2 data with a P$_{\mu}$>70$\%$, and they found two BSS candidates in this cluster. The difference in the above-mentioned catalogs could be due to the adopted age criteria, selection method, and different membership probability cut-offs used in the two studies. \\

We obtained the MESA Isochrones \& Stellar Tracks (MIST) for the UVIT and \textit{Gaia} EDR3 filters from an updated MIST online database\footnote{\url{https://waps.cfa.harvard.edu/MIST/interp_isos.html}} to identify and classify distinct evolutionary sequences in the cluster \citep{2016ApJ...823..102C, 2018ApJS..234...34P}. We considered isochrones with \big[$\alpha$/Fe\big] = +0.0, metallicity, Z = 0.017210 \citep{2021MNRAS.502.4350S}, not incorporating initial rotation. 
Cluster parameters such as age, extinction, and distance modulus, adopted to fit the isochrone to the observed optical CMD, are 775 Myr, A$_{V}$=0.6\,mag, and (m$-$M)$_{V}$=12.56, respectively \citep{2021MNRAS.502.4350S}. The overplotted isochrone (solid green line) over the observed optical CMD is displayed in Figure~\ref{optcmd}. We notice that the isochrone appears well-matched to the observed CMD along the main-sequence, sub-giant branch (SGB), but it is not reproducing the observed position of the red clump. To account for this mismatch along the red clump, \citep{2018MNRAS.480.3739B} suggested that there might be a problem in the calibration of the models for the red clump or the conversion between theoretical properties of the isochrones (temperature, gravity, and luminosity) to observational space in \textit{Gaia} filters is off.\\
We also selected the YSSs based on their location in the optical CMD, as they have colors in between the turn-off (TO) and RGB and appear brighter than the SGB. We have chosen two such stars marked with yellow colored filled symbols shown in Figure~\ref{optcmd}.

\subsection{FUV-optical CMDs}
\label{sec:fuvcmds}
This section presents the FUV-optical CMDs generated by cross-identifying common stars between optical and our FUV detections. We cross-matched the sources detected in the UVIT FUV filters with the \textit{Gaia} EDR3 with a maximum separation of 1\farcs3, which is the typical FWHM of the PSF for the UVIT filters. 
To plot the FUV-optical CMDs, first, we made the magnitude system adopted by \textit{Gaia} similar to that of UVIT. That is, we transformed the Vega magnitude system used in the \textit{Gaia} photometric system to the AB system using the photometric zero points reported in the \textit{Gaia} EDR3 documentation\footnote{\url{https://gea.esac.esa.int/archive/documentation}}. 

\begin{figure*}[!htb]
        \centering
        \includegraphics[height=7.5cm]{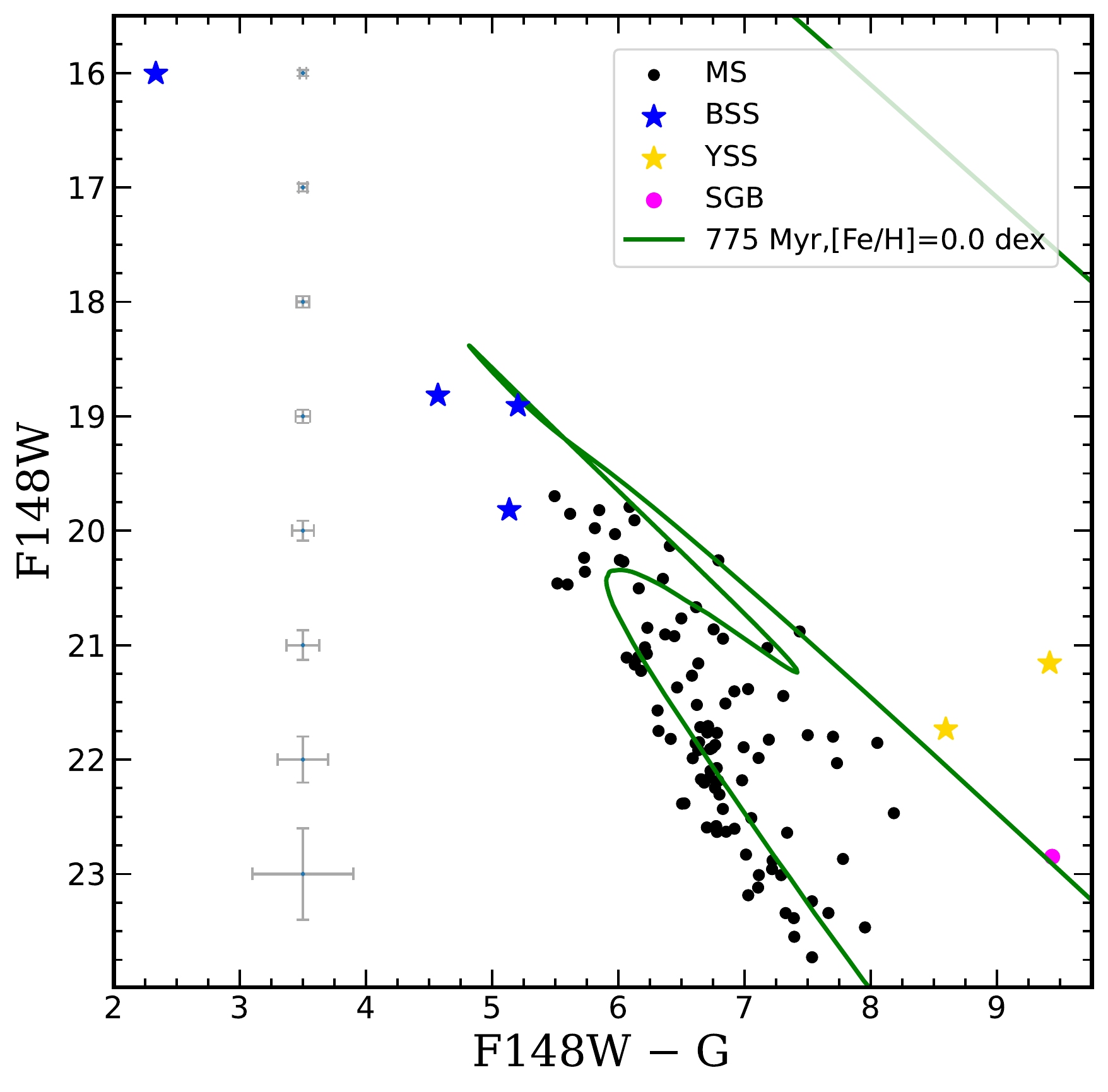}
        \includegraphics[height=7.5cm]{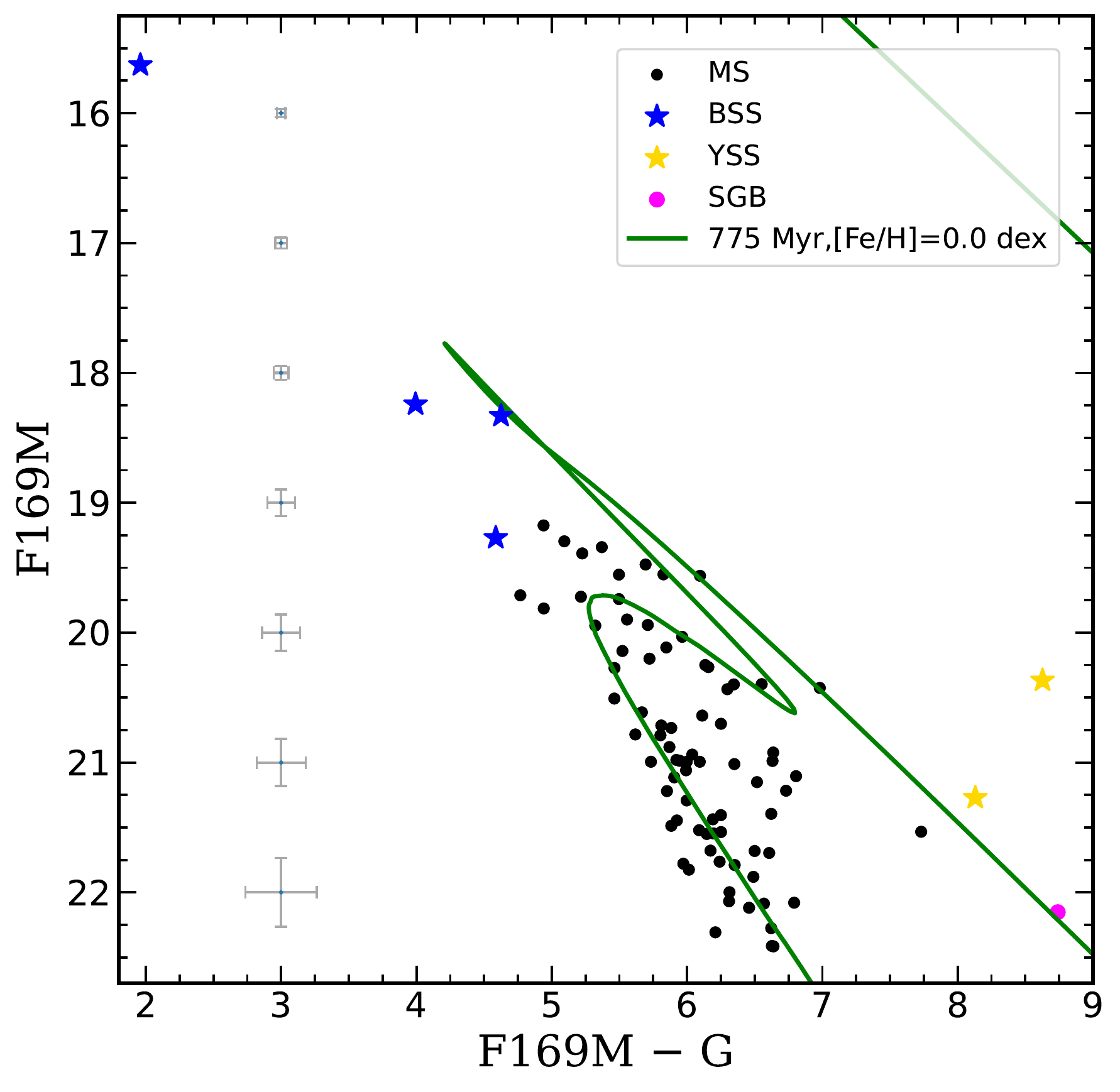}
	\caption{FUV-optical CMDs using F148W and F169M passbands of NGC\,2818 of confirmed members cross-identified using UVIT FUV and \textit{Gaia} EDR3 catalog. The error bars (median) are shown in gray color on the left side of each panel. The rest of the details are the same as in Figure~\ref{optcmd}.}
    \label{fuvcmds}
\end{figure*}

We have created and shown the FUV-optical CMDs for cluster members in Figure~\ref{fuvcmds} using F148W and F169M filters. We note that a similar trend of detected stellar populations is observed in the other two filters (F154W \& F172M). The error bars displayed in all FUV CMDs are estimated as the median of the stars' errors at a chosen magnitude range.
The FUV-optical CMDs are also over-plotted with updated MIST isochrones \citep{2016ApJ...823..102C} to compare the locations of the distinct sequences predicted by the theoretical models with the observed ones. In all FUV images, hot and bright stars such as BSSs, YSSs, and MS are detected. We have detected 4 BSSs out of 5 previously known in the literature \citep{2021A&A...650A..67R}. Four detected BSSs are confirmed RV and PM members. Two YSSs are also identified in all FUV images. We note that these stars are well-separated and brighter than the theoretical isochrone presenting the SGB sequence in all FUV-optical CMDs, in turn confirming their classification as YSSs. RGB and Red clump stars are too faint to be detected in the FUV.\\
The FUV-optical CMDs show a large scatter along MS, as shown in Figure~\ref{fuvcmds}, unlike optical CMD. The overlaid isochrones in all FUV-optical CMDs help to trace the MS scatter. We note that a few MS stars are brighter than theoretical MSTO not reproduced by isochrones. These might have high rotational velocities accounting for this feature. Some of them may be binaries or potential BSSs. 
 One BSS is found to be very hot and bright in all FUV-optical CMDs compared to the other three BSSs. This BSS can be an exciting candidate to characterize, as it might have a hot WD companion. As two YSSs are detected in all FUV images and found to be bright in all FUV-optical CMDs, these stars also might have a hot companion, which leads to their detection in the FUV images. These are intriguing targets further to understand their formation and evolution in the clusters.

\subsection{Extended MS turn-off in FUV CMDs }
\label{rot_models}
\begin{figure}[!htb]
  \hspace*{-0.5cm} 
        \includegraphics[width=1.1\columnwidth]{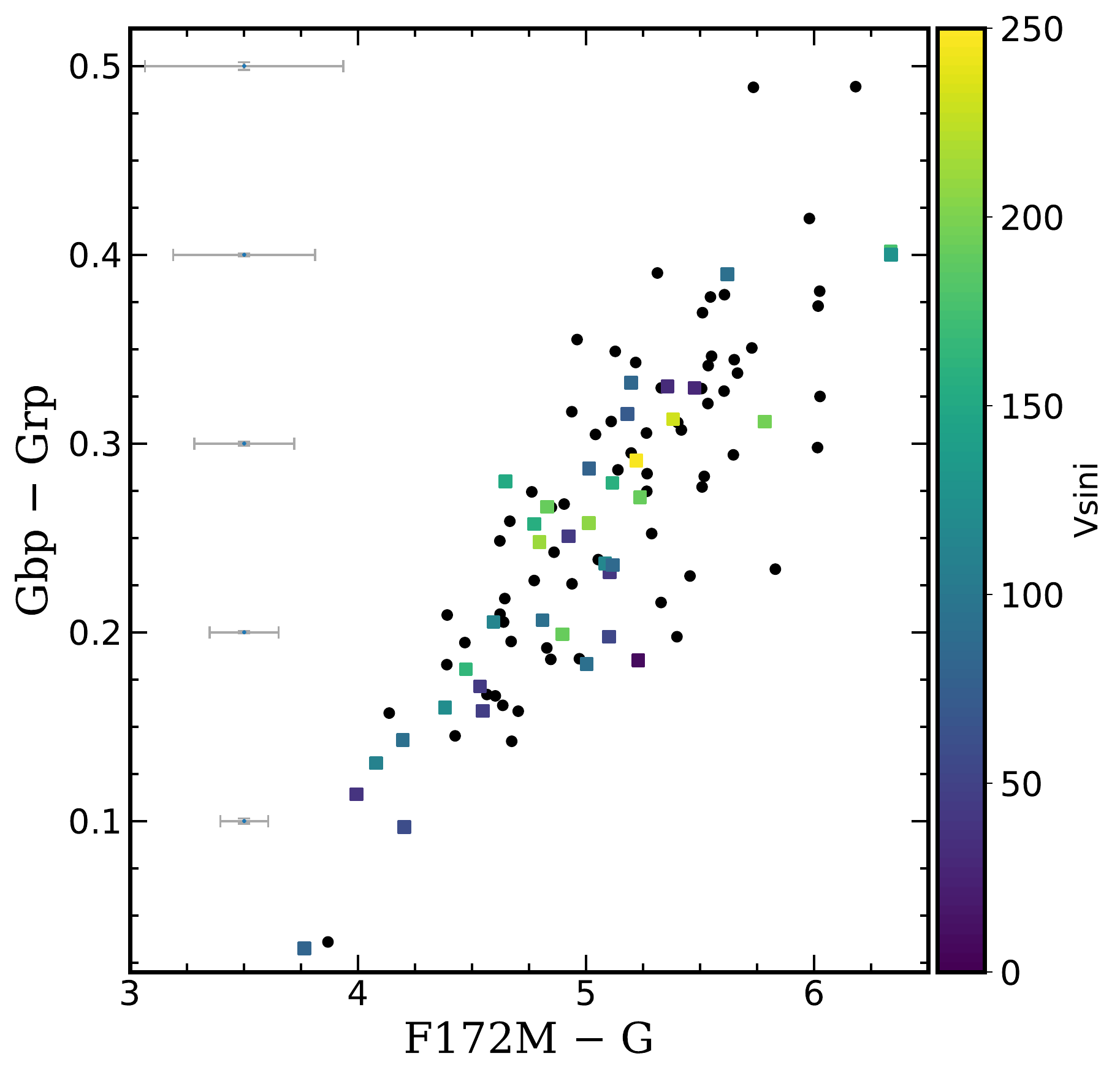}
	\caption{F172M$-$G vs. Gbp$-$Grp color-color plot of all stars detected with UVIT color-coded by their measured Vsini values. Stars with black color symbols do not have estimated Vsini values.}
    \label{colorplot}
\end{figure}

In order to check the sensitivity of UVIT colors to the $T_{eff}$ affected by the rotational velocity, we plot (Gbp$-$Grp) vs. (F172M$-$G) color as shown in Figure~\ref{colorplot}, which indicates a linear relation. The range of \textit{Gaia} color is only 0.4 mag whereas F172M$-$G spans about 3.0 mag, which makes F172M$-$G color more sensitive and responsive to rotational velocity. F172M$-$G color is preferred over F169M$-$G because the band F172M allows only continuum light, and no chromospheric or transitional emission lines are seen in late-type stars in FUV.

\begin{figure*}[!htb]
        \centering
        \includegraphics[height=6cm]{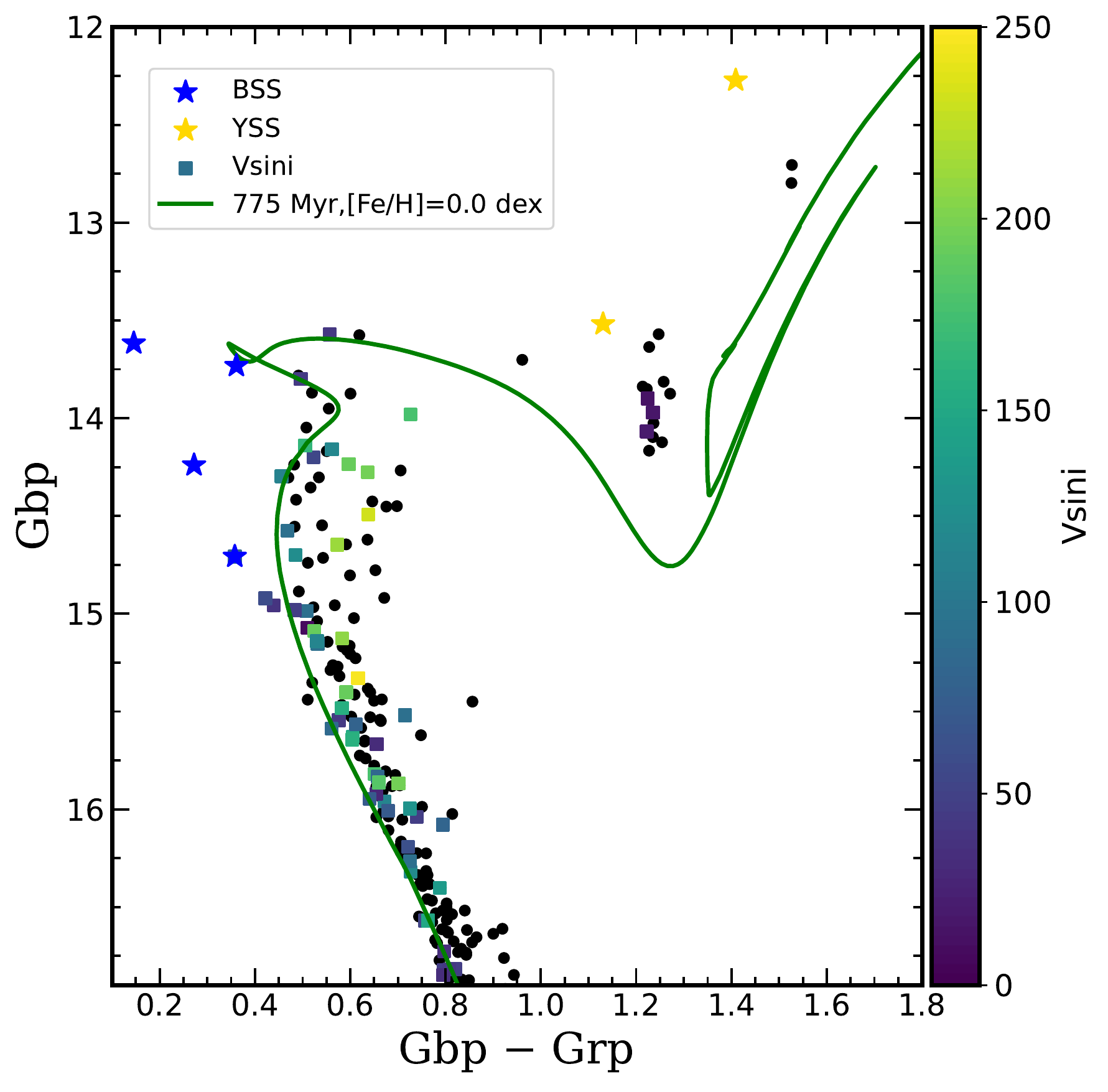}
        \includegraphics[height=6cm]{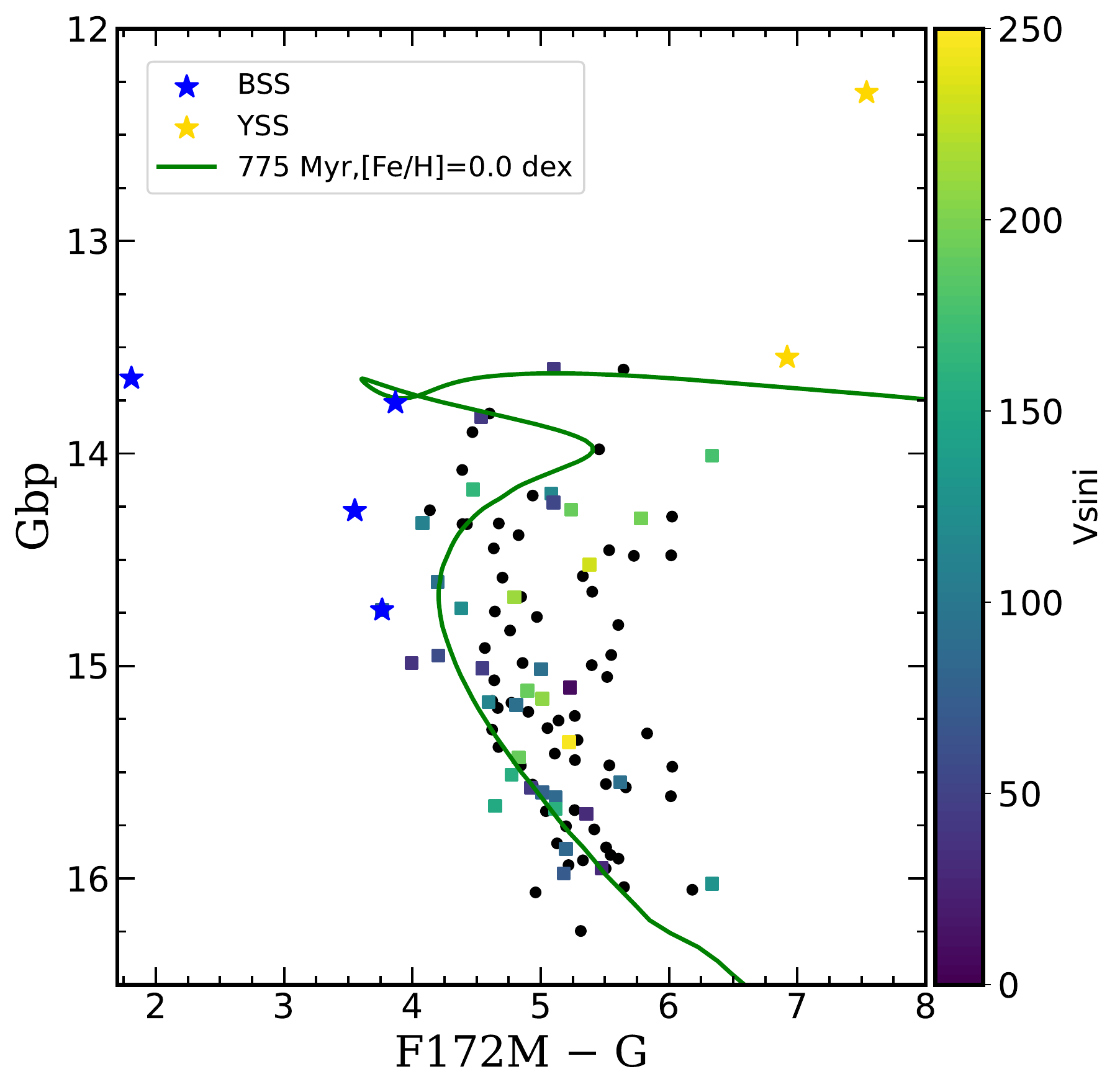}
        \includegraphics[height=6cm]{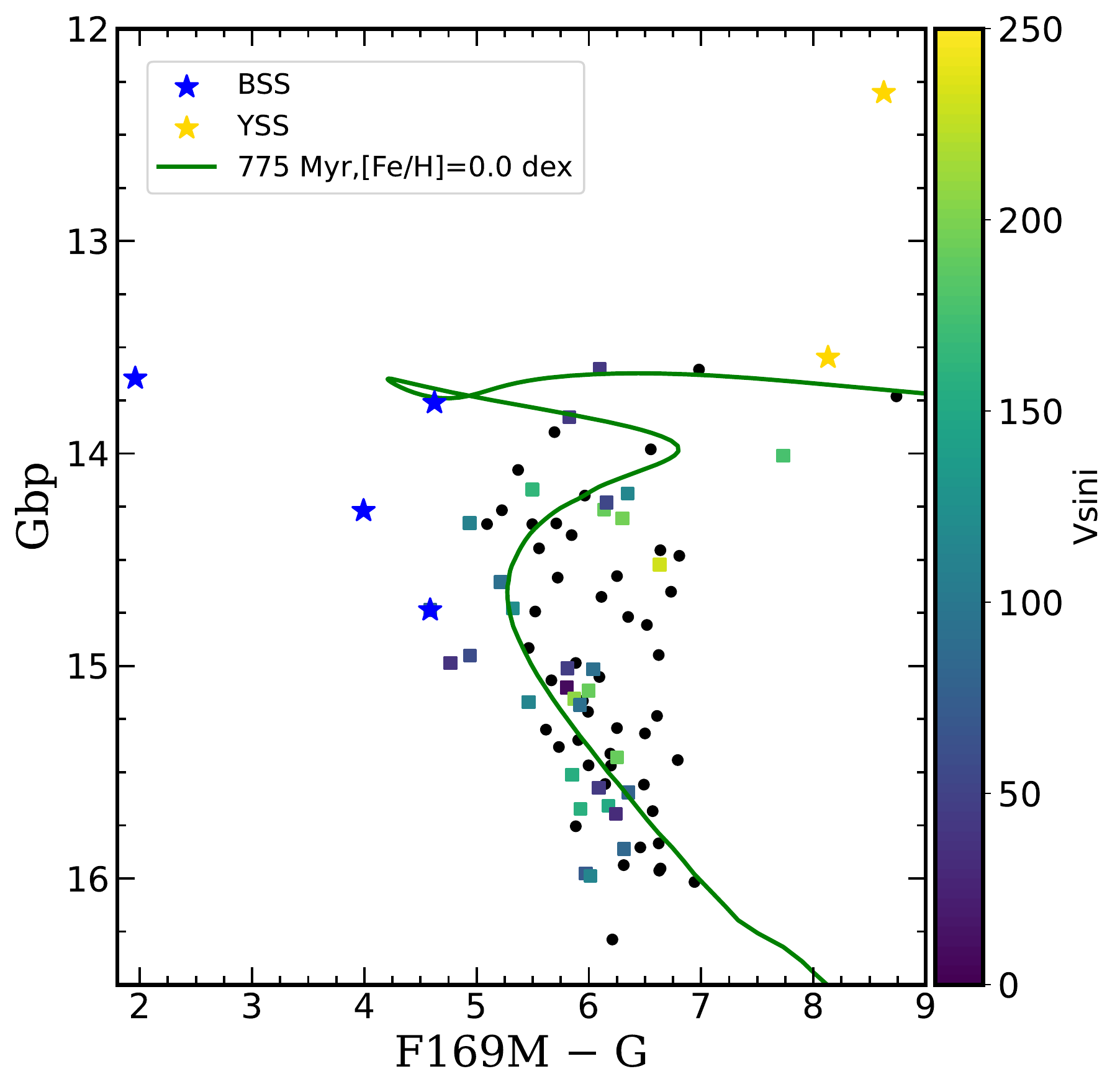} 
        \includegraphics[height=6cm]{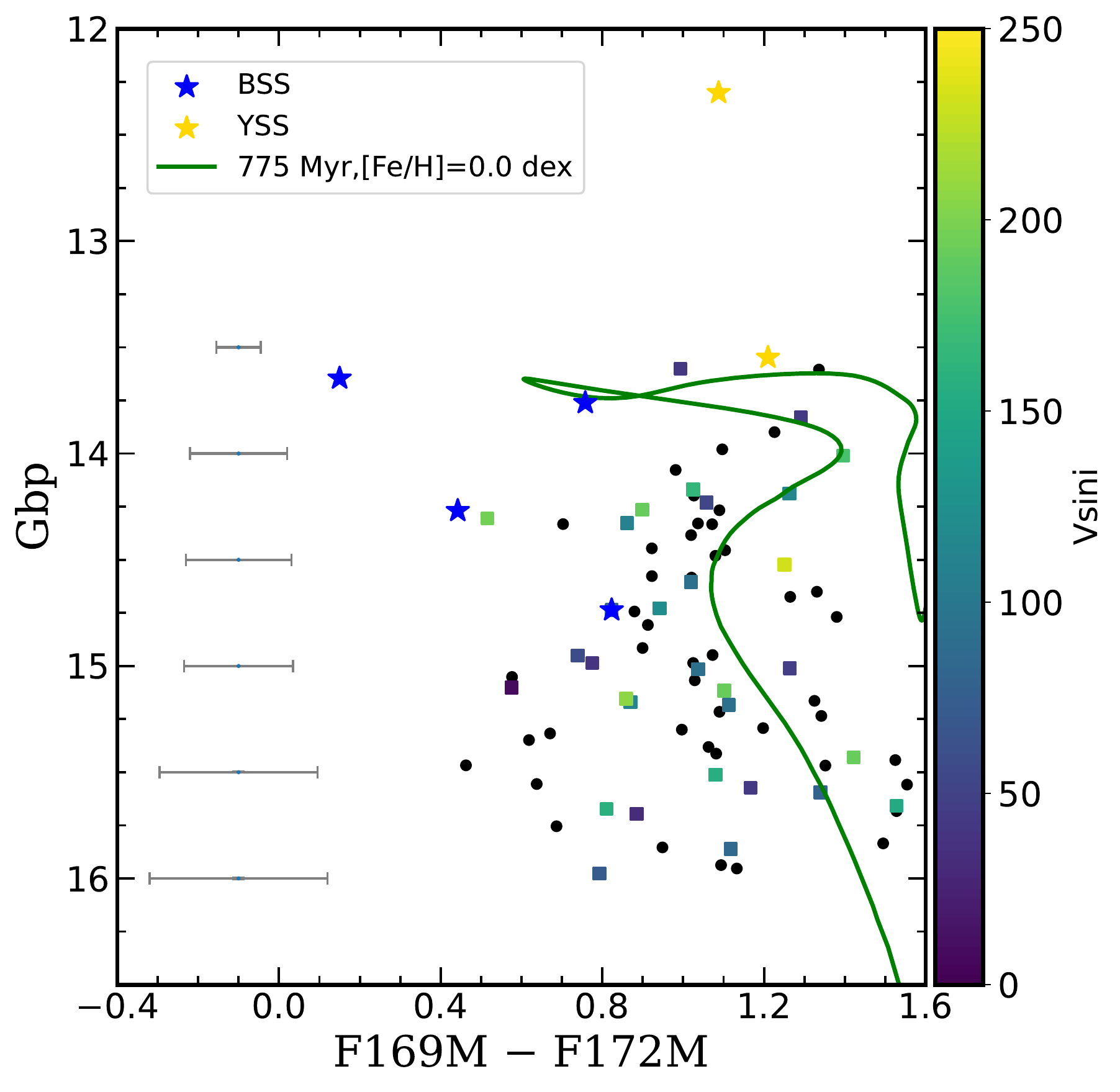}
	\caption{Optical (upper left), F172M-G vs Gbp (upper right), F169M-G vs. Gbp (lower left), and F169M-F172M vs. Gbp (lower right) CMDs of NGC\,2818 members color-coded by measured Vsini values. The rest of the details are the same as in Figure~\ref{optcmd}.}
    \label{extmsto}
\end{figure*}

Comparison of the CMD, F172M$-$G vs. Gbp (Fig.~\ref{extmsto} upper right) with CMD of Gbp$-$Grp vs. Gbp (Fig.~\ref{extmsto} upper left) shows the sensitivity of F172M$-$G color. The bend in the isochrone in F172M$-$G vs. Gbp CMD at a color of 4.0 indicates the beginning of eMSTO prominently (unlike Fig.~\ref{extmsto}, left panel), and all the stars right of the isochrone show high rotational velocity.
The MS comprises stars with both high and low rotational velocities. However, the CMD of F169M$-$G vs. Gbp exhibits some more aspects. From the comparison of F169M$-$G color with F172M$-$G in Fig.~\ref{extmsto}, we find that the former is redder than the latter. It can be due to the fact that the F169M flux in late-type stars is smaller than at F172M. Moreover, the predicted colors using the theoretical isochrones are following the same trend.\\
It is well known that MS stars later than about F2 would possess coronal and transitional regions as evidenced in the FUV region by emission lines of C IV, He II, Si IV, N V, N IV, etc. \citep{1979ApJ...229L..27L, 1987ASSL..129..259J}. Prominent lines like C IV and He II occur in the F169M band region (unlike the F172M band). The F154W and F148W would contain a few more emission lines in addition to C IV and He II. Thus, the CMD of F169M−G vs. Gbp shows that the MS stars are shifted bluewards to the isochrone, probably suggesting the presence of transitional region lines. Even in the F169M$−$F172M vs. Gbp CMD shown in the lower right panel of Figure~\ref{extmsto}, it is evident that most stars have bluer colors than the theoretically expected ones from isochrones. It is to be noted that all stars on the blue edge of the MS in CMD of F169M$−$G vs. Gbp (15<Gbp<16, 5<F169M$-$G<6) show high rotational velocity in contrast to CMD of F172M$-$G vs. Gbp (15<Gbp<16, 4<G172M$-$M<5). It is fairly well established that high rotational velocities enhance the coronal and transitional line emissions \citep{1981ApJ...248..279P, 2020ApJ...902....3L}. Thus, it is consistent with the suggestion that high rotation stars are on the blue side because of high emission line activity in total contrast to the MS of F172M$−$G vs. Gbp CMD. This phenomenon sets into stars redder than (Gbp$-$Grp) $\sim$0.5 mag.\\

\begin{figure*}[!htb]
\centering
\includegraphics[width=0.92\columnwidth]{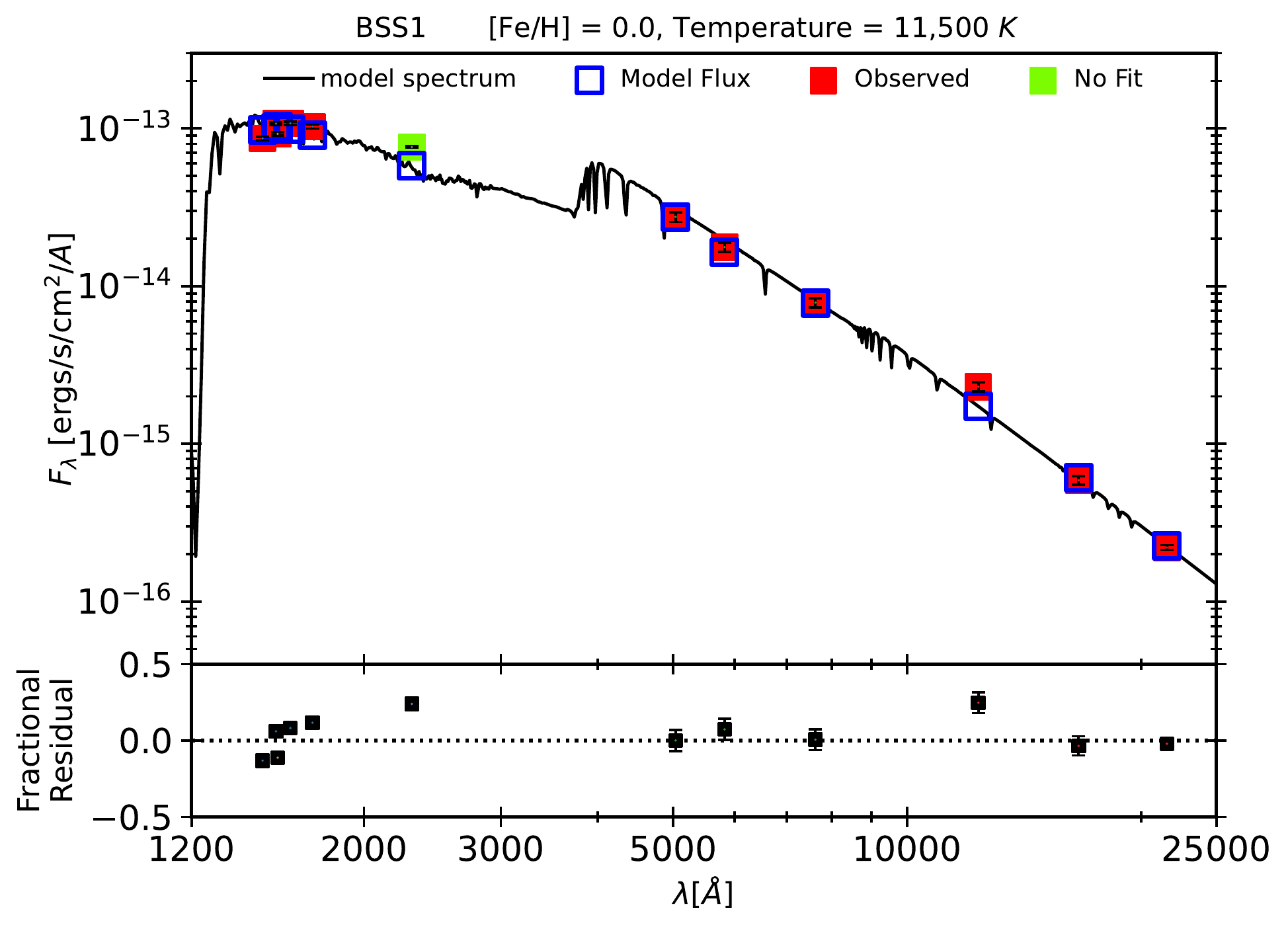}
\includegraphics[width=0.89\columnwidth]{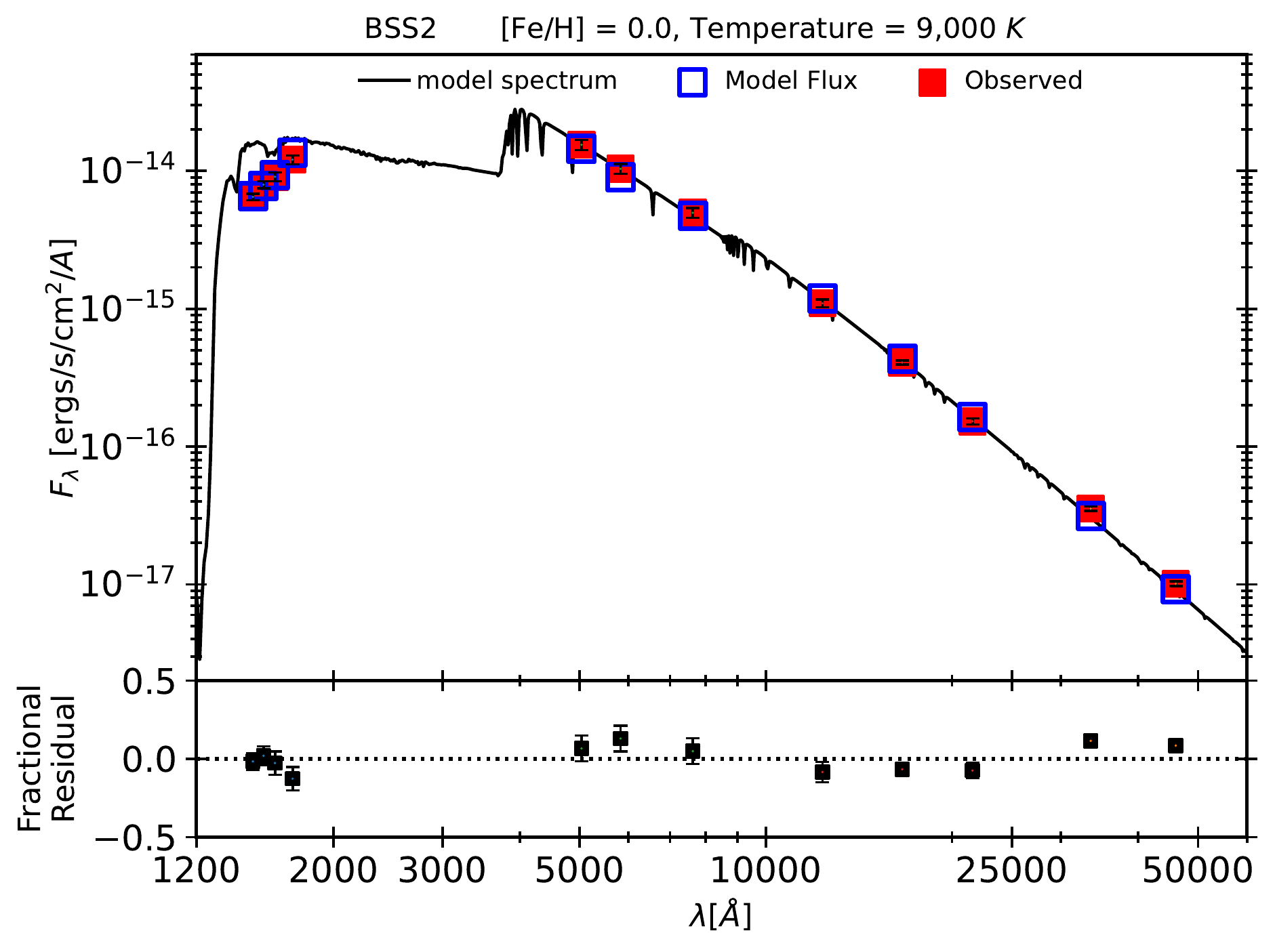}
\includegraphics[width=0.89\columnwidth]{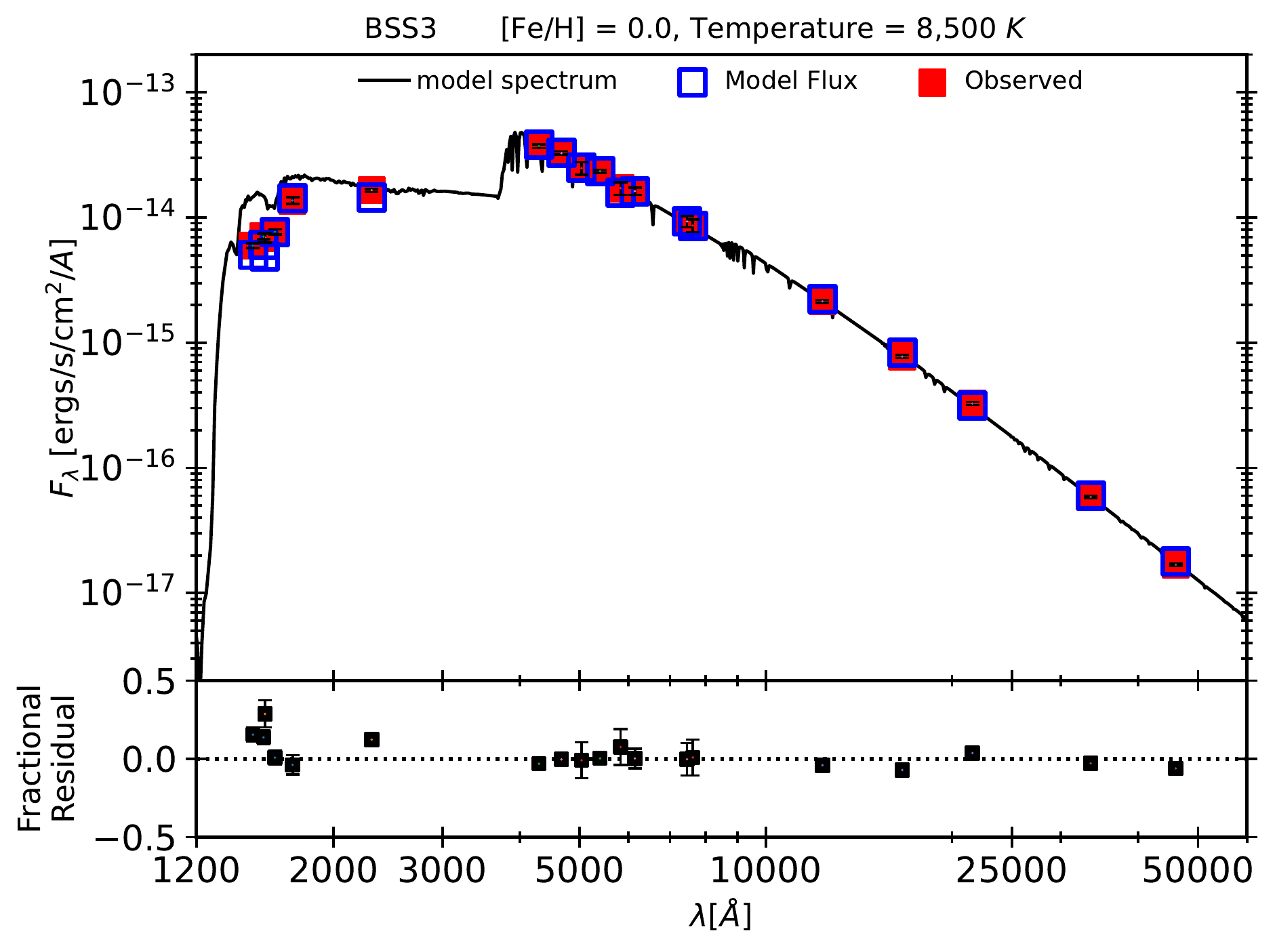}
\includegraphics[width=0.89\columnwidth]{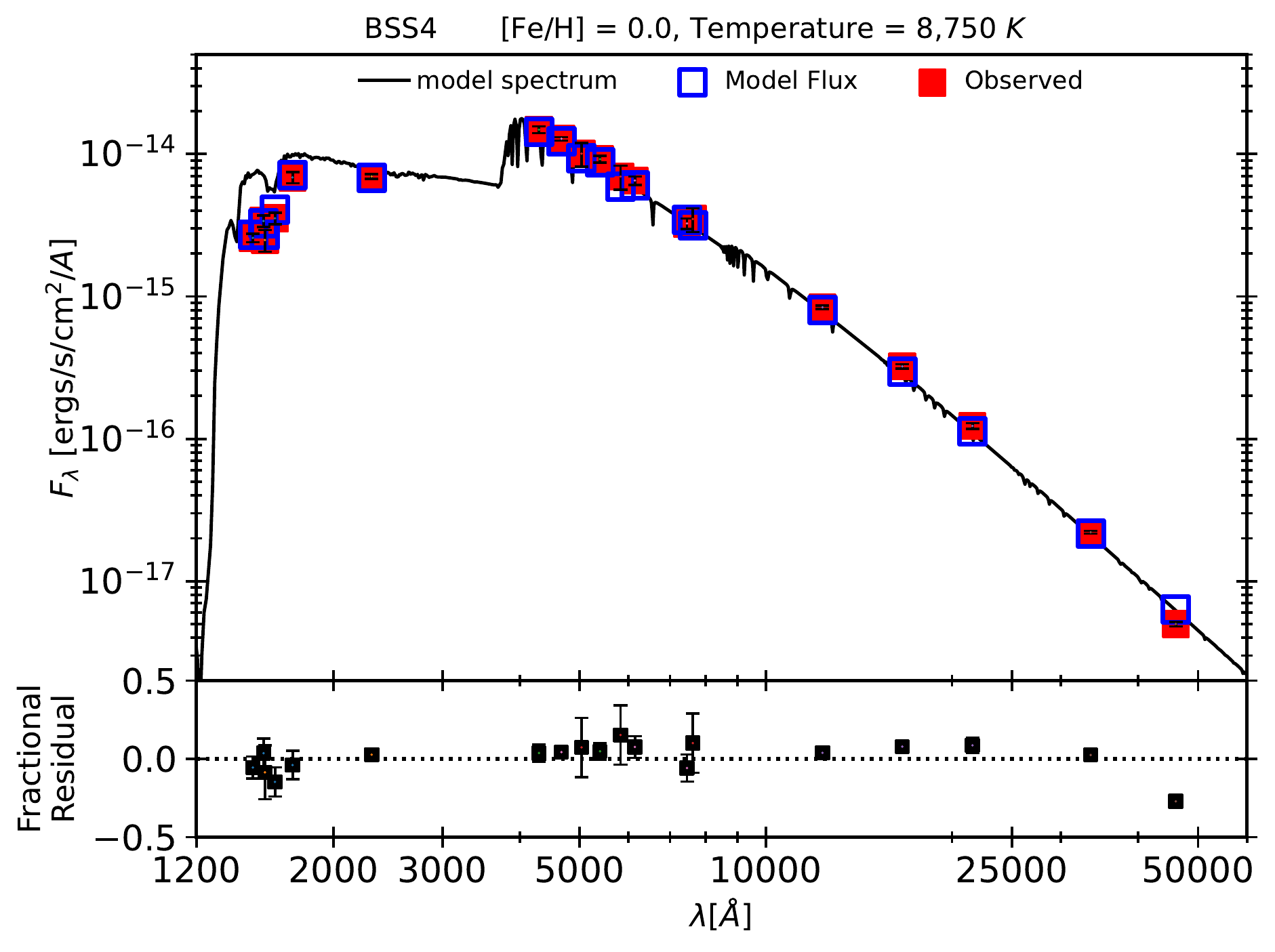}
 \caption{SEDs of four BSSs detected with UVIT. Extinction correction has been incorporated in all the observed photometric fluxes from UV to IR. The BSS ID adopted in this work is shown in each figure. The gray color presents the best-fitting Kurucz model spectrum in all the plots. The data points that are excluded in the SED fit are shown with the orange color-filled symbol. The bottom panel in all the SEDs illustrates the residual between the observed fluxes and model predictions.}
 \label{bsseds}
\end{figure*}

\begin{figure*}[!htb]
\centering
\includegraphics[width=2\columnwidth]{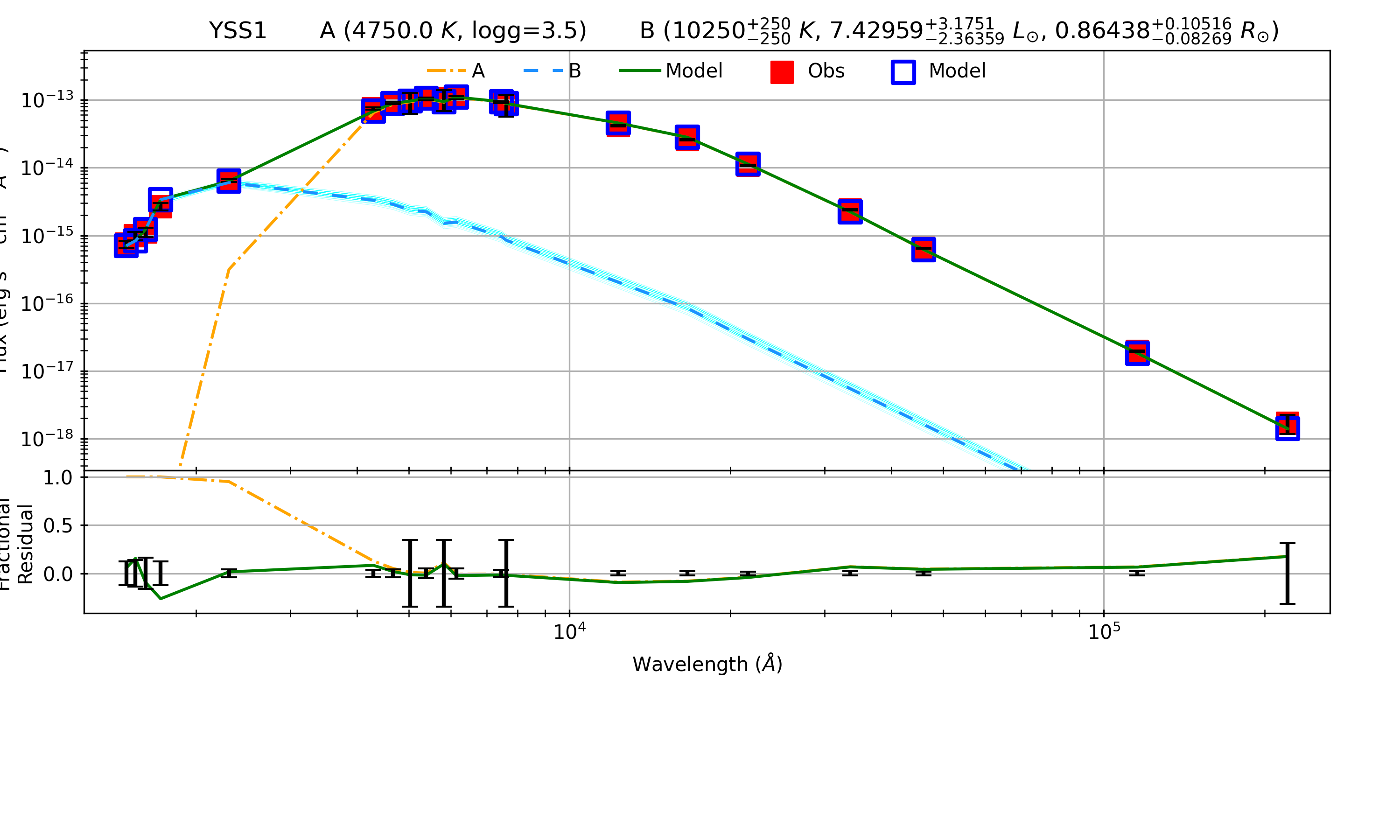}
\includegraphics[width=2\columnwidth]{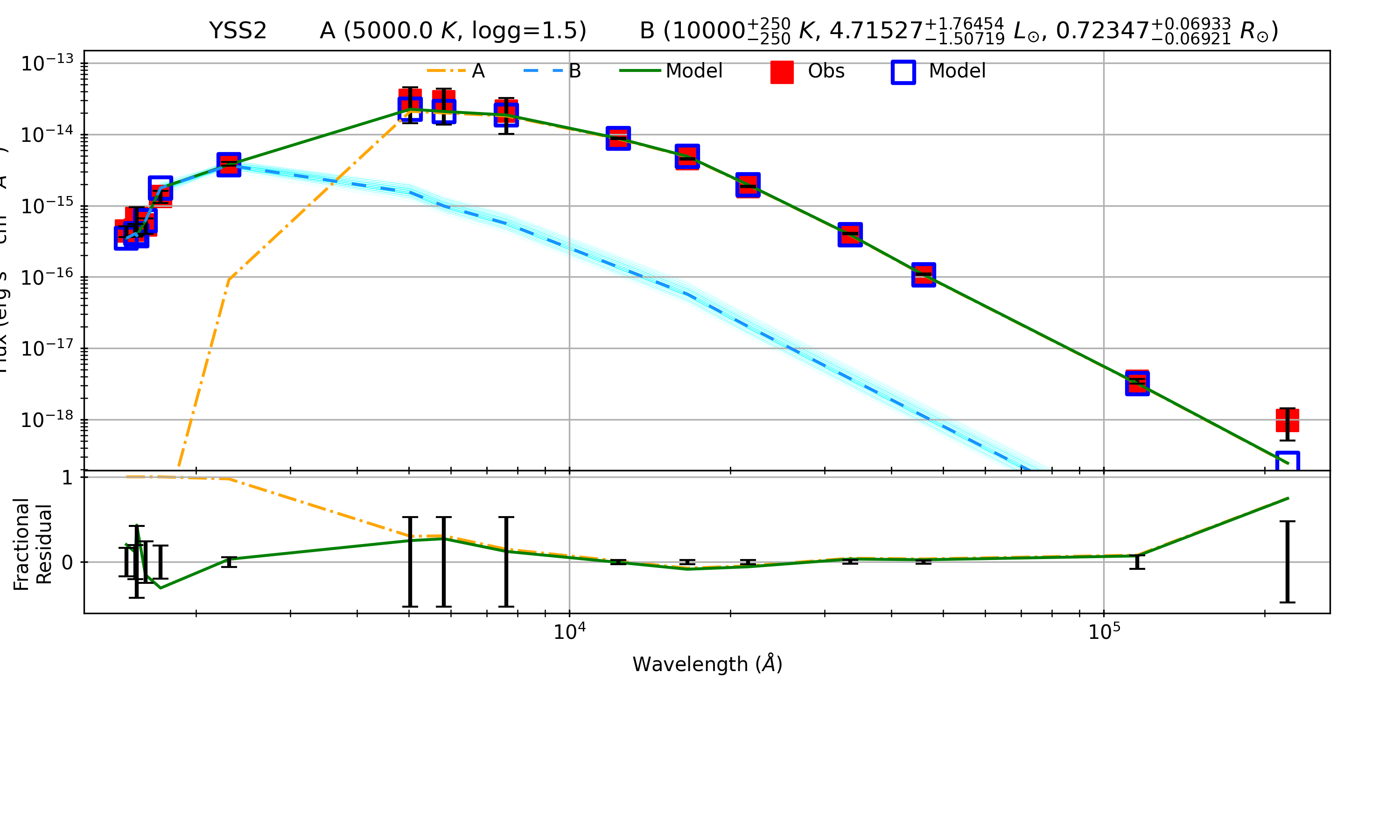}
 \caption{Double-fit SEDs of YSSs. The meaning of all the symbols is displayed in the legend. The star IDs and parameters of two components obtained from the fit are shown on the top of both SEDs. The green color represents the composite model flux along with the observed fluxes marked with red symbols. Orange dotted-dash and blue dashed lines indicate Kurucz and Koester models used to fit the star's cooler and hotter components, respectively. The middle panel presents the fractional residual (Orange dashed line) corresponding to the single fit as well as the composite fit (Green solid line). The fractional observational uncertainties in the flux are also shown here. The values of $\chi_{red}^{2}$ and modified $\chi_{red}^{2}$ parameter, namely $vgf_{b}^{2}$ representing the best-fit are displayed in the lower panel.}
 \label{ysseds}
\end{figure*}

\begin{figure*}[!htb]
\centering
\includegraphics[width=1.01\columnwidth]{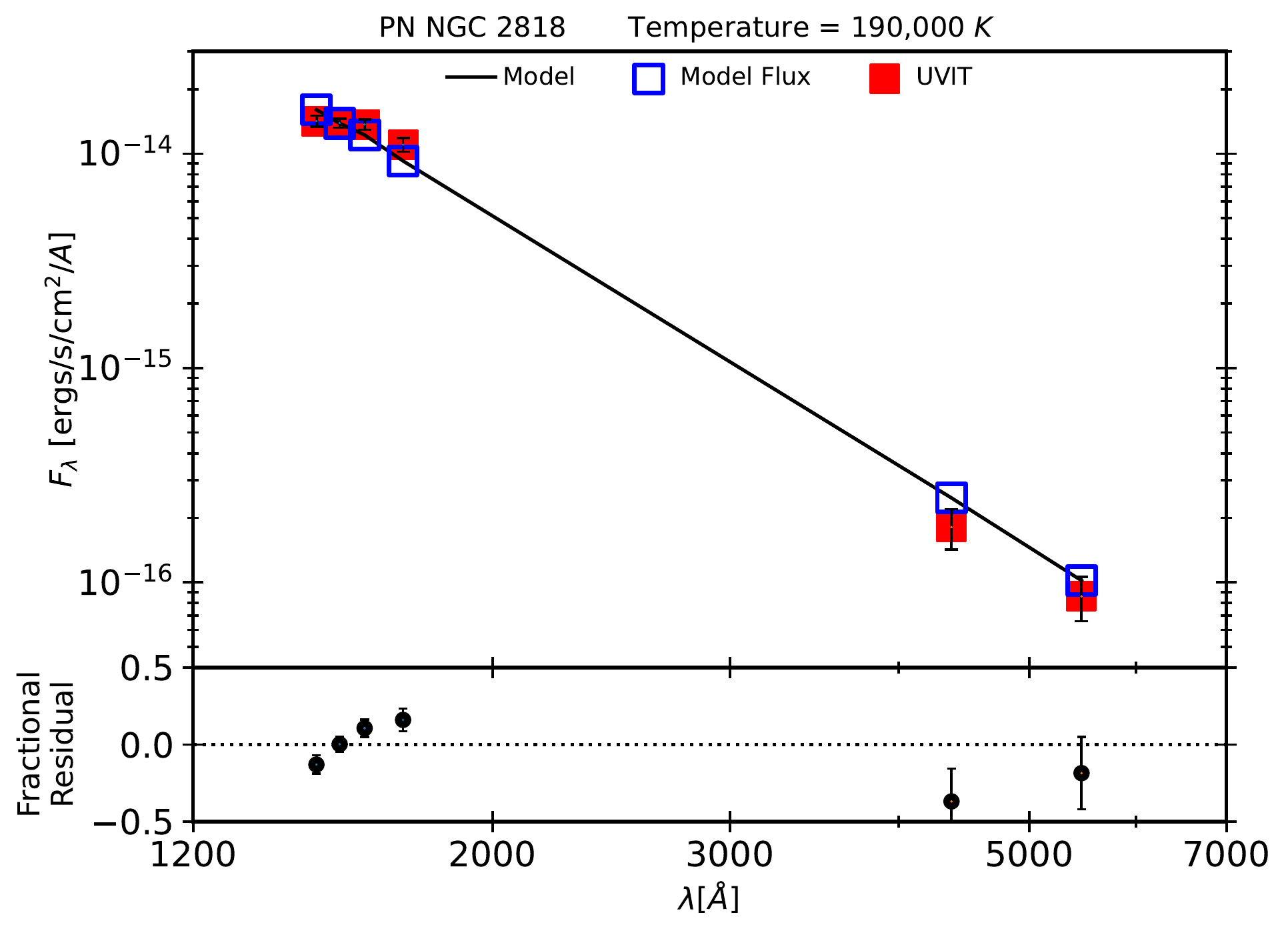}
\includegraphics[width=0.97\columnwidth]{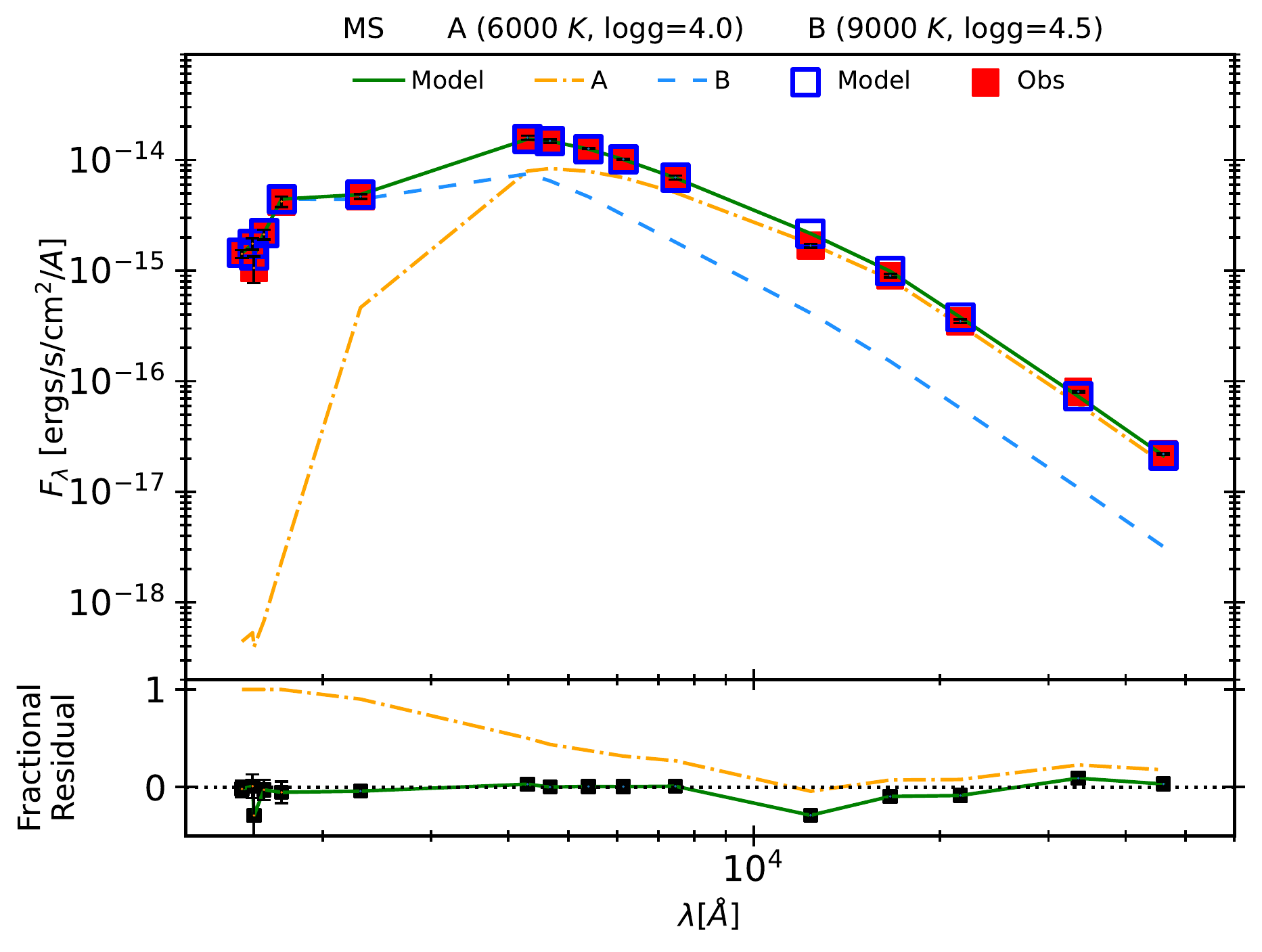}
 \caption{SED fit of the CSPN (left panel) and MS star (right panel) after taking into account the  extinction correction. The black solid line represents the theoretical TMAP model fit to the observed fluxes shown with red symbols. The best-fit $T_{eff}$ value is displayed in the figure. The rest of the details are the same as in Figure~\ref{bsseds} and \ref{ysseds}.}
 \label{PNesed}
\end{figure*}

\begin{table*}
    \begin{center}
    \large
     \caption{Stellar parameters obtained from best SED fit of BSSs detected with UVIT in NGC\,2818. Column 1 lists the star ID used in the paper. Columns 2 and 3 display the RA and DEC of all the stars considered for fitting, respectively. The {\it $T_{eff}$}, luminosities, and radii of all-stars, along with errors, are tabulated in columns 4, 5, and 6, respectively. Columns 7 and 8 lists the reduced $\chi^{2}$ value corresponding to the best fit and ratio of the number of photometric data points ($\frac{N_{fit}}{N_{tot}}$) used for the fit to the total number of available data points.}
    \label{tab2}
	\begin{tabular}{ccccccccccc} 
		\hline
		\hline
		 Star ID & RA (deg) & DEC (deg) & {\it $T_{eff}$} (K) & $\frac{L}{L_{\odot}}$  & $\frac{R}{R_{\odot}}$ & ${\chi}_{red}^2$ & Vgf & $Vgf_{b}$ & $\frac{N_{fit}}{N_{tot}}$\\
		\hline
         BSS1 & 139.0306 & -36.59184 & $11,500 \pm 250 $ & $91.55 \pm 17.54$ & $2.39 \pm 0.22$ & 12.9 & 12.9& 1.53 & 11/12\\
         BSS2 &	139.0279 & -36.59178 & $9,000 \pm 250$ & $32.99	\pm	6.31$ &	$2.31 \pm 0.21$ & 3.1 & 3.1 & 0.88 & 12/12\\
         BSS3 & 139.1633 & -36.43083 & $8,500^{+500}_{-250}$ & $52.28	\pm	9.84$ &	$3.30 \pm 0.31$ & 4.8 & 4.8 & 1 & 19/19\\
         BSS4 &	139.0276 & -36.6423 & $8,750 \pm 250$ & $20.97 \pm 3.94$ & $1.94 \pm 0.18$ & 4.9 & 4.9 & 0.91 & 19/19\\
	\hline
	\end{tabular}
	 \end{center}
\end{table*}

\begin{table*}
  	\begin{center}
	\small
	\caption{Derived parameters of YS and MS stars from the composite SED fit. The different models used to fit the cooler (A) and hotter (B) components of the SEDs are presented in column 5. The rest of the columns have the same meaning as depicted in Table~\ref{tab2}.}
	\label{tab3}
\begin{tabular}{cccccccccccc} 
		\hline
		\hline
		 Star ID & RA (deg) & Dec (deg) & Type &  Model Used & {\it $T_{eff}$} (K) & $\frac{L}{L_{\odot}}$  & $\frac{R}{R_{\odot}}$ & ${\chi}_{red}^2$ & Vgf & $Vgf_{b}$ & $\frac{N_{fit}}{N_{tot}}$ \\
		\hline
		YSS1 & 139.0523 & -36.57946 & A & Kurucz & $4,750 \pm 125$ & $338.1 \pm 63.25$ & $27.01 \pm 2.49$ & 5.6 & 5.6 & 0.36 & 20/20 \\ 
		 &  &  & B & Koester & $10,250 \pm 250$ & $ 7.43^{+3.17}_{-2.36}$ & $0.864^{+0.105}_{-0.083}$ & 4.3 & 4.3 & 0.61 & \\ 
		YSS2 & 138.9976 & -36.58243 & A & Kurucz & $5,000 \pm 250$ & $78.91 \pm 15.55$ & $10.93 \pm 1$ & 3.5 & 3.5 & 0.71 & 16/16 \\ 
		 &  &  & B & Koester & $10,000 \pm 250$ & $4.72^{+1.76}_{-1.51}$ & $0.723^{+0.069}_{-0.069}$ & 2.4 & 2.4 & 0.81 & \\ 
		 MS & 139.0592 & -36.60989 & A & Kurucz & $6,000 \pm 125$ & $18.35 \pm 3.47$ & $3.98 \pm 0.37 $ & 7.3 & 7.2 & 0.99 & 18/18 \\ 
		 &  &  & B & Kurucz & $9,000 \pm 125$ & $10.79 \pm	2.04$ & $1.36 \pm 0.125$ & 7.3 & 7.2 & 0.99 & \\ 
		 
		\hline
	\end{tabular}
	\end{center}
\end{table*} 

\begin{table*}
    \begin{center}
    \small
     \caption{Derived parameters of PN NGC\,2818 from the best SED fit. The notation of all columns is the same as described in Table~\ref{tab2}}
    \label{tab4}
	\begin{tabular}{ccccccccccc} 
		\hline
		\hline
		 Star ID & RA & DEC & Model Used & {\it $T_{eff}$} & $\frac{L}{L_{\odot}}$  & $\frac{R}{R_{\odot}}$ & ${\chi}_{red}^2$ & Vgf & $Vgf_{b}$ & $\frac{N_{fit}}{N_{tot}}$\\
		 & (deg) & (deg) & & (K) & & & & & &\\
		\hline
		 PN NGC\,2818 & 139.0061 & -36.62707 & TMAP(Grid3) & $190,000 \pm 8080.40$ & $826.75 \pm 225.21$ &	$0.026 \pm 0.002$ & 8.3 & 8.3 & 4.5 & 6/6\\
    \hline
	\end{tabular}
	 \end{center}
\end{table*}

\section{Spectral Energy Distribution Fits}
\label{sec:SEDs}
It is well demonstrated in previous studies of exotic stellar populations, such as BSSs in OCs, that they are products of stellar interactions. There might be a chance of detecting a binary companion in the case of BSSs and YSSs. SEDs of such systems can be used to obtain the parameters of the multiple components. In this section, we present the multi-wavelength SEDs constructed for the BSSs, YSSs, and CSPN identified with UVIT to derive their atmospheric parameters like effective temperature ({\it $T_{eff}$}), luminosity (L), and radius (R). We aim to probe the physical nature of these stars and probable hot companions, if present, by estimating their stellar parameters and placing them on the HR diagram. SEDs are generated with the observed photometric data points spanning a wavelength range from FUV-to-IR and fitted with selected theoretical models. We made use of the virtual observatory tool, VOSA (VO Sed analyzer, \citealp{2008A&A...492..277B}) for SED analysis. The details of the SED fitting technique are described in \cite{Rani2021}. In addition to $\chi_{red}^{2}$, VOSA calculates two extra parameters, Vgf and $Vgf_{b}$, known as modified $\chi_{red}^{2}$ to estimate the goodness of fit in case the observational flux errors are too small. The value of $Vgf_{b}$ should be less than 15 to achieve a reliable SED fit \citep{2021MNRAS.506.5201R}.

The Kurucz stellar atmospheric models are employed to create synthetic SEDs \citep{1997A&A...318..841C, 2003IAUS..210P.A20C} for BSSs and YSSs, which have observed photometric data points covering a wavelength range from UV to IR. The free parameters available in the Kurucz model are {\it $T_{eff}$}, metallicity, and log\,{\it g}. To fit the observed SEDs of the stars, as mentioned earlier with Kurucz models, we assumed {\it $T_{eff}$}, and log\,{\it g} as free parameters, and fixed the value of metallicity \big[Fe/H\big] = 0.0, close to the cluster metallicity. We adopted the range of {\it $T_{eff}$} from 5,000-50,000 $K$ and log\,{\it g} from 3.5-5 dex in the Kurucz models. We combined the photometric data points of UVIT (4 passbands) with \textit{GALEX} (2 passbands), \textit{Gaia} EDR3 (3 passbands) \citep{2018A&A...616A..12G}, SDSS (3 passbands), APASS (2 passbands), 2MASS (3 passbands), and WISE (4 passbands) to generate the observed SEDs. VOSA makes use of Fitzpatrick reddening law \citep{1999PASP..111...63F, 2005ApJ...619..931I} to compute the extinction in different passbands and correct for extinction in observed fluxes for the provided $A_{V}$. VOSA utilizes the Markov chain Monte Carlo (MCMC) approach to estimate the uncertainties in the stellar atmospheric parameters obtained using the SED fit. We estimated the radius (R) of the star using the scaling relation $M_{d} = \big(\frac{R}{D}\big)^2$, where D is the distance to the cluster and  $M_{d}$ is the scaling factor.

We conducted SED fitting analysis for four BSSs, two YSSs, and PN, as described in the following subsections. 

\subsection{Blue Straggler Stars}
\label{sec:BSSEDs}
The best-fitted SEDs for all BSSs are shown in Figure~\ref{bsseds}, where the lower panel of each SED depicts the fractional residual between the observed and predicted fluxes. The overplotted black solid line presents the synthetic Kurucz model spectrum created using the parameters corresponding to the best-fit SED. The star IDs adopted in this work are displayed on top of each SED. We observe that the SEDs of all BSSs are seemed to be well-fitted with a single model, as the residual is close to zero in all SEDs. Since the observed flux errors are very small for all the filters used, the error bars (shown with black color) are smaller than the data points. We list their parameters corresponding to the best fit in Table~\ref{tab2}. We obtain $Vgf_{b}$ values for all BSSs to be around 1, indicating the good SED fits, and all the derived fundamental parameters are also reliable. The BSSs have a $T_{eff}$ range of 8,500$-$11,500 $K$, and radii of 1.9$-$3.3 $R_{\odot}$. Now, here arises the two possibilities about the nature of these stars: 1) either all BSSs are single stars, 2) or they are binaries with a very faint companion, not able to detect by the UVIT observations. If these stars are single, they are likely to be formed via the merger of the component stars in a  binary.


\subsection{Yellow Straggler Stars}
\label{sec:YSSEDs}
Figure~\ref{ysseds} presents the SEDs of two stars classified as YSSs in this work. In this figure, the lower panel represents the fractional residual, i.e., the ratio of the difference between the observed and model flux ($F_{obs} - F_{model}$) and the observed flux at every given data point.
We can see in Figure~\ref{ysseds} that both YSSs are showing significant UV excess as a single model could not fit the entire SED. It can also be noticed in the fractional residual plot showing a rise in flux in the UV wavelengths for a single spectrum fit (shown as an orange dash-dotted line in the figure). To fit the hotter component of the system, first, we gave excess for wavelength less than 3000 {\AA} and fitted the cooler component that includes the optical and IR data points with the Kurucz model by selecting {\it $T_{eff}$} range from 3,500$-$50,000 $K$ and log{\it g} from 1.5$-$2.5 dex. From the single fit, the computed values of {\it $T_{eff}$} of the YSS1 and YSS2 are 4,750 $K$ and 5,000 $K$, respectively. The radius of YSS1 and YSS2 is 27 R$_{\odot}$ and $\sim$ 11 R$_{\odot}$, respectively. From their temperature and radii, we infer that they are in the giant phase of stellar evolution. After obtaining the stellar parameters of the cooler component, then we used Binary SED Fitting\footnote{\url{https://github.com/jikrant3/Binary_SED_Fitting}} code to fit the hotter part of the SED. The full details of this code are well described in \cite{2021JApA...42...89J}. As we expect the hotter component to be compact, we have used the Koester WD model \citep{2009ApJ...696.1755T, 2010MmSAI..81..921K}. In this model, the range of free parameters {\it $T_{eff}$} and log{\it g} is 5,000$-$80,000 $K$ and 6.5$-$9.5, respectively. The double fit of both stars is shown in Figure~\ref{ysseds}, where the Kurucz model fit is shown with an orange dash-dotted line, and the Koester model fit with a light-blue dashed line. The composite fit is marked with a solid green line. The fractional residual in both plots is close to zero for all observed data points indicating how well the double component fit reproduces the observed SED. This is even evident from the $vgf_{b}$ values (close to 1) computed from the SED fitting of both stars. The estimated parameters of both YSSs from the best binary fit are tabulated in Table~\ref{tab3}. 
From the double fit, we estimate the {\it $T_{eff}$} of the hotter companion of YSS1 and YSS2 are 10,250 $K$ and 10,000 $K$, respectively. The values of parameters such as {\it $T_{eff}$}, luminosities, and radii of the stars are mentioned on the top of each SED. 

\subsection{PN NGC~2818}
\label{nebsed}
 As we have shown in the previous section, the PN NGC\,2818 most likely has a physical association with the cluster; it will be interesting to characterize its central star to obtain information about its progenitor.
 We can clearly see the CSPN in the FUV image, as shown in Figure~\ref{fuvimage},  implying its very high temperature. The magnitude of CSPN is a vital parameter to study its evolution as it can be used to determine the stellar parameters. The magnitude of the CSPN in optical filters was measured by \cite{1988A&A...197..266G}. As CSPN is well observed in all FUV images, therefore we have calculated the magnitude of the central star by performing the PSF photometry on the FUV images acquired in 1st and 2nd epoch observations. We have subtracted the nebular background in assessing the magnitude of the CSPN.
The external extinction and distance to the nebula are considered to be the same as that of the cluster. Four FUV UVIT data points are combined with two optical photometric data points from \cite{1988A&A...197..266G} to construct the observed SED of the nebula. As the central star seemed to be very hot, we have fitted its SED with the T\"ubingen NLTE Model Atmosphere Package (TMAP) (Grid3) model used for hot stars \citep{2003ASPC..288..103R, 2003ASPC..288...31W}. This model grid spans a range of atmospheric parameters such as $50,000 K \leq T_{eﬀ} \leq 190,000 K$, $5.0 \leq log{\it g} \leq 9.0$, and $0 \leq X_{H} \leq 1$.  It is important to note that we took into account the external extinction while fitting its SED but did not incorporate the internal extinction in the nebula. We have noticed that {\it $T_{eff}$} derived using the TMAP model fit to the observed SED corresponds to their upper limit, which indicates that this star is likely to be hotter than the estimated temperature from this model. The stellar parameters computed from the best-fit SED of the nebula are summarised in Table~\ref{tab4}.\\

\subsection{MS stars}
\label{MS_stars}
We also have constructed the SEDs for the MS stars
detected with UVIT, for which rotational velocity information was available in the literature to investigate their nature. Apart from that, we also have considered the MS stars for SED analysis for which rotational velocity was not estimated earlier, and their position in all FUV-optical CMDs was not matched with their expected one. 31 MS stars with the known rotational velocity are identified with UVIT in two epochs. Other than these stars, 6 MS stars are brighter than MS turn-off in FUV CMDs. We have used the Kurucz models to fit their observed SEDs to obtain their physical parameters and check their binarity. Out of 37 stars, we observed that only one MS star shows significant FUV excess, as displayed in the right panel of Figure~\ref{PNesed}, whereas other stars show less or mild UV excess that could not be fitted with a double component SED. Chromospheric activity in the above star cannot account for UV excess as it is exceptionally high compared to the model. The other possibility to explain this excess is the presence of a hot companion that mainly emits at shorter wavelengths. To account for the presence of the hot companion, we fitted the entire SED with the Kurucz model using the binary fit task from VOSA. The double component fit for this star is found to be satisfactory (Right panel of Figure~\ref{PNesed}), and the best-fit parameters computed are tabulated in Table~\ref{tab3}. The radii of both components suggest that they are not quite on the MS. The cooler companion is likely to be a sub-giant ($R/R_{\odot} \sim$ 4.0), whereas the hot companion has a smaller radius ($R/R_{\odot} \sim$ 1.36) when compared to the MS star of similar temperature  ($R/R_{\odot} \sim$ 6.0). It might be possible that this is a post-mass transfer system where the hotter component is the donor, and the cooler component is still bloated after gaining mass. The rotational velocity (Vsini) of this star is around 39 km/s.\\ 


\section{Evolutionary Status}
\label{sec:status}
\begin{figure*}[!htb]
\hspace{-0.3cm}
\centering
\makebox[\textwidth]
{
\includegraphics[width=0.98\textwidth]{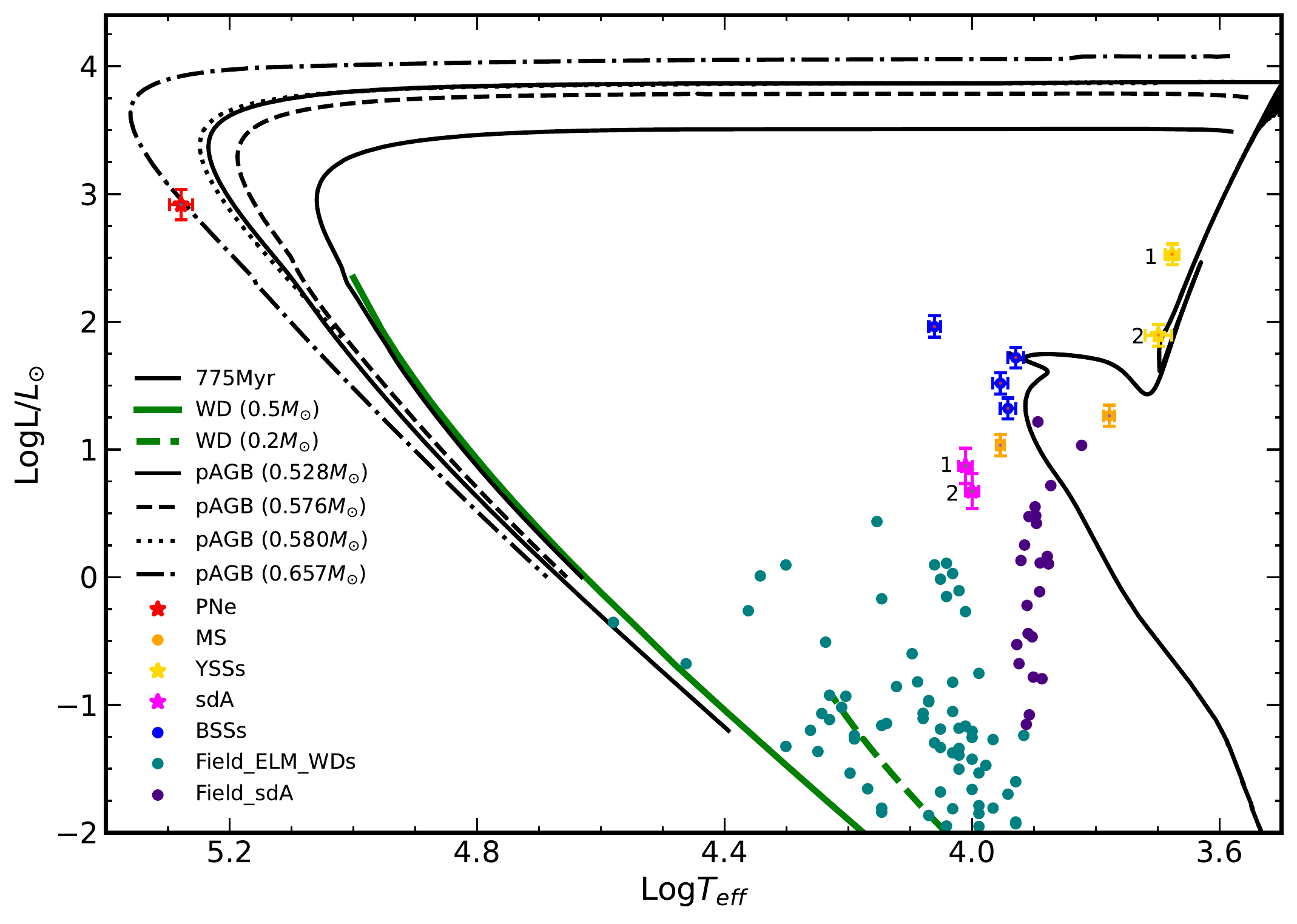}
}
\caption{HR diagram of the bright stars identified with UVIT. Various evolutionary tracks are presented from the beginning of the MS to the moment when a star has entered to the  stage, followed by the WD cooling sequences. All these tracks are generated for cluster metallicity and age. The pAGB sequences with different final masses are shown here to compare the location of the CSPN marked with a red star symbol. BSSs and YSSs are displayed with blue-filled circles and yellow star symbols, respectively. The hotter companions of YSSs are shown with magenta star symbols. In addition, Field ELM WDs and A-type subdwarfs represented with cyan and purple symbols are also placed in the HR diagram to compare the position of the hot companions of both YSSs. Green color solid and dashed lines correspond to the DA-WD tracks with different masses.}
\label{bastievotrack}
\end{figure*}

Placing the stars on the HR diagram provides information about their evolutionary stage and helps in probing the nature of the hot companions in the case of binary stars. To examine the evolutionary status of exotic stars considered in this study, we have plotted the theoretical evolutionary sequences starting from the MS to the moment the star has entered the tip of the RGB stage. These tracks are taken from MIST models computed by \cite{2016ApJ...823..102C, 2018ApJS..234...34P} and selected for the cluster age and metallicity close to the cluster metallicity. The stellar parameters estimated from the single SED fit for four BSSs are plotted in the HR diagram. The meaning of the color and symbols are marked in Figure~\ref{bastievotrack}. We can notice in Figure~\ref{bastievotrack} that BSSs are lying bluer to the MS track, suggesting that these four stars belong to the BS evolutionary phase.

The location of two YSSs on the HR diagram is near the theoretical RGB sequence. It indicates that their progenitors' (BSSs) have already evolved into a giant phase where the contracting helium core is surrounded by the hydrogen-burning shell. The hot companions of both YSSs seemed to be compact in nature, as indicated by their estimated radii suggesting they might belong to the WD or extremely low mass (ELM) WD or subdwarf stage of stellar evolution. In addition to the MS tracks, we have presented the DA-type WD cooling sequences with masses 0.5$M_{\odot}$ and 0.2$M_{\odot}$ taken from \cite{2011ApJ...730..128T} in Figure~\ref{bastievotrack}. From comparing the position of the hot companions of both YSSs with theoretical WD cooling tracks, we notice that their location is not reproduced by them, implying that they still have not entered the WD stage. While there are non-DA type WDs that are believed to result from mergers, they are not expected to be found in OCs because the merger process would take longer than the age of the cluster. 


In order to find out where ELM WDs fall in the HR diagram, we have used the field ELM WD catalog provided by \cite{2016ApJ...818..155B}. They have estimated the $T\rm{_{eff}}$ and log\,{\it g} values of the considered ELM WD sample in their paper. To place them on the $T_{eff}$ vs. luminosity plot, SED fitting technique is used to estimate the luminosity of all ELM WDs (Priv. Comm. Vikrant Jadhav). The extinction correction has been incorporated in all the stars. All field ELM WDs are marked as cyan-filled symbols in Figure~\ref{bastievotrack}. We note that the hot companions of the YSSs are more luminous than the field ELMs with a similar temperature. 

As the location of the binary companions of YSSs is not reproduced by the WD tracks as well as ELM WDs, we further suspect that they might belong to the class of A-type subdwarfs (sdA) as they are lying near the general location of subdwarfs in the HR diagram. sdA stars are supposed to occupy the location between the dwarfs and WDs in the HR diagram; hence, they are more compact than dwarfs, indicating a higher log\,{\it g} value. \cite{2017ApJ...839...23B} performed a detailed study of sdA stars to investigate their physical nature and a possible link to the ELM WDs. We used the field sdA catalog to locate their positions on the HR diagram. As only effective temperatures of all sdA stars were available in the catalog, we used the SED fitting technique to determine their luminosities. The extinction in the visual band ($A_{V}$) for these stars was estimated using the reddening map provided by \cite{2011ApJ...737..103S}. We have taken care of the extinction correction in the observed fluxes in different bands of all sdA stars. The distances to these stars are available in the \textit{Gaia} EDR3 catalog. We have used the distances reported in \cite{2021yCat.1352....0B}, estimated using \textit{Gaia} EDR3 catalog, and they all fall within a range of $\sim$1.5 to 8 kpc. The sdA stars are displayed with purple-filled symbols in the HR diagram. The hot companions of YSSs are found to be hotter than the similarly luminous field sdAs and more luminous than the similarly hot field sdAs. 

From this comparison, we suggest that they are most likely to be sdA stars
formed through a binary mass transfer scenario. These binaries are probably a post-mass-transfer system consisting of an A-type subdwarf candidate and a YS star.
We also checked the position of the hotter and cooler components of the MS star on the HR diagram displayed with orange-color symbols. The hotter component occupies a location bluer than theoretical isochrone, might be evolving to the sdA type star, whereas the cooler component occupies the location expected for sub-giants. The evolution of this star might be similar to the YSS as the cooler component is evolving to the giant stage, whereas the hotter component later might end up as sdA. Thus, we speculate that this system might be a progenitor of the YSSs detected in this cluster.


 Further, we have used the pAGB models computed by \cite{2016A&A...588A..25M}  to deduce the evolutionary state of the CSPN. We adopted the cluster metallicity (Z=$\sim0.02$ dex) to select the pAGB tracks. Tracks with a range of final mass as shown in Figure~\ref{bastievotrack} are presented from the beginning of the pAGB phase when the H-rich envelope drops below $M_{env} = 0.01M_{*}$  to the moment the star has already entered its WD cooling sequence at $L_{*} = L_{sun}$. The estimated parameters of the PN from the SED fit are plotted in the HR diagram (Red filled symbol). From the comparison to these theoretical pAGB tracks, we observe that CSPN is found to be located on the track (Black dash-dotted line) corresponding to the final mass  $0.657M_{sun}$. It can be noted from here that the star has already entered the WD cooling phase.
 
\section{Discussion}
\label{sec:dis}
We have conducted an observational study of OC NGC\,2818 and the PN within its field using FUV medium-resolution space-based imaging data from UVIT aboard \textit{AstroSat}.
This paper aims to use the most accurate and complete \textit{Gaia} EDR3 data on stellar astrometry and photometry in the nearby intermediate age OC NGC\,2818 to establish the membership probability of known stars and to deduce the evolutionary state of exotic stars. Since the stars reside in the central area of the cluster, we have confined ourselves with the consideration of the inner part of the cluster with a radius of 30$\arcmin$ and selected 37508 stars brighter than G = 21 mag. Using the GMM method to pick out the PM members, we have chosen 718 stars as the cluster members with $P_{\mu} > 50 \%$ and considered them further to identify their FUV counterparts with UVIT. FUV-optical and FUV CMDs were generated for the cluster members and overlaid with the MIST isochrones to compare the position of different observed evolutionary sequences with theoretically expected ones. MIST isochrones are found to match well with the observed sequences in FUV-optical CMDs, but in FUV CMDs, especially F169M$-$F172M vs. Gbp, most of the detected stars in both filters are lying blueward of their expected location from isochrones. 

In all FUV images, we have identified four BSSs, two YSSs, and MS based on their location in the optical as well as FUV-optical CMDs. Then, we performed the SED analysis to deduce their physical properties to evaluate their nature. The $T_{eff}$ of BSSs estimated from SED fit ranges from  8500$-$11500 $K$, hinting that they are quite hot, consistent with the young age (700$-$800 Myr) of the cluster. In the previous studies of BSSs in other OCs conducted using UVIT data, the $T_{eff}$ range varies from cluster to cluster depending upon its age. The temperature range of BSSs in OC M67 (4 Gyr) is 6250$-$9000 $K$ \citep{2019ApJ...886...13J}, in King\,2 (6 Gyr) 5750$-$8500 $K$ \citep{2021JApA...42...89J}, in OC NGC\,188 (7 Gyr) 6100$-$6800 $K$ \citep{2015ApJ...814..163G}. In intermediate-age OCs such as NGC\,7789 (1.6 Gyr) \citep{2022MNRAS.511.2274V} and NGC\,2506 (2.2 Gyr) \citep{2022MNRAS.516.5318P}, BSSs span a temperature range from 7250$-$10250 $K$, and 7750$-$9750 $K$, respectively. The SEDs of all BSSs are well-fitted with a single model, and we suggest that collisions leading to the mergers might explain their formation in this cluster. Another plausible possibility is that they might have a faint WD companion undetectable with UVIT. If this is the case, then the second prominent scenario to explain their existence in star clusters, i.e., mass transfer in close binaries, will dominate over the previous one. Moreover, mass transfer in binaries will dominate in OCs as they are less dense and compact than GC systems. Further, spectroscopic analysis of these stars will help to confirm their nature.\\

Two YSSs, from their SED fits, are found to be binaries, and the location of YSSs and their hot components in the HR diagram suggests that cool components are already in the RGB phase. In contrast, hot components most plausibly belong to sdA class. We infer from here that these two stars are post-mass-transfer systems where BSS (accretor) has evolved into a giant stage and became YSS, and the donor star into a sdA. In addition, a spectroscopic study performed by \cite{2001A&A...375...30M} of RGB stars, including these two stars, found that they are spectroscopic binaries, confirming our result. Their radial velocities estimated by them also verify their membership. Hence, we suggest that these two stars to be formed via a mass transfer scenario in the cluster.\\

From the comparison of the distance, extinction, RV and
PM values of the PN with the cluster, it turns out that it is a most likely member of the cluster. \cite{2003RMxAA..39..149B, 2008ApJ...674..954B} estimated the $T_{eff}$ from the ionization modeling of the nebula as $T_{eff}$ 149,000 $K$ and log\,{\it g} of 7.1 (however, this might also be dependent on the distance assumed). \cite{2016MNRAS.459..841M} gives the $T_{eff}$ as 160,000 $K$. \cite{1988A&A...197..266G} estimate the HI Zanstra temp 175,000K and HeII Zanstra temp of 215,000K. \cite{1986A&A...162..232K} derived the luminosity ($L_{*}=851L_{\odot}$) and radius ($R_{*}=0.038R_{\odot}$) of CSPN using optical observations, and adopting the identical distance to the nebula as that of the cluster (d=3.5 kpc). The atmospheric parameters of CSPN determined using the SED fitting technique are more or less in agreement with the previous estimations. Based on the comparison of the central star's location with the predicted ones from the theoretical  models in the HR diagram, the central star's mass turns out to be 0.66 $M_{\odot}$. \cite{2018ApJ...866...21C} presented the WD initial–ﬁnal mass relation (IFMR) for progenitor stars of $M_{initial}$ from 0.85 to 7.5 $M_{\odot}$. In their Figure 5, they displayed the comparison of the Initial–Final Mass Relation (IFMR) estimated for the observed sample with the theoretical isochrones. For a WD with a mass of 0.66 $M_{\odot}$, the initial mass of the progenitor is estimated to be $\sim$2.1 $M_{\odot}$ (From their Fig. 5). In this work, the MSTO mass of this cluster determined using isochrone fit is $\sim$2 $M_{\odot}$. The previously reported turn-off mass for this cluster and the initial mass of the nebula's progenitor are $\sim$2.1 $M_{\odot}$, and $2.2\pm0.3$ $M_{\odot}$, respectively \citep{1984ApJ...287..341D}. Our estimations are consistent with the previous ones. From the comparison of the cluster turn-off mass and progenitor mass, we infer that PN is quite likely a cluster member. Thus, this study showcases the significance of using the FUV data to study the exotic populations and late stages of the evolution of intermediate-mass stars in OCs.

\section{Summary and Conclusions}
\label{sec:summary}
The main results from this work can be summarized
as follows:

\begin{itemize}
    \item In this study, we employed UVIT observations onboard \textit{AstroSat} to identify BSSs and YSSs in the open cluster NGC\,2818, and also characterize the CSPN. We further created the optical and UV-optical CMDs of member stars co-detected using UVIT and \textit{Gaia} EDR3 data in this cluster.
    \item The PM members of the cluster are obtained using \textit{Gaia} EDR3 data, and we found that PN NGC\,2818 might be a member of this cluster, consistent with the previous studies.
    \item  As this cluster is young, hot and bright stars such as BSSs, YSSs, and MS are detected in all FUV images.
    \item  To compare the observations with theoretical predictions, optical and UV-optical CMDs are overlaid with non-rotating MIST isochrones generated for respective UVIT and \textit{Gaia} filters. The theoretical isochrones reproduce the features of all CMDs quite well.
    \item  The FUV-optical CMDs prominently show the eMSTO phenomenon already reported in this cluster, consistent with the previous studies.
    \item We characterized the four detected BSSs in the cluster, and a single model fits well to all the observed SEDs. We suggest from the single model fits that these stars might have a faint WD companion that could not be detected with UVIT’s detection limit or result from the merger of two close binaries.
    \item We suggest the presence of two YSSs in this cluster based on their location in the CMDs. Both YSSs were found to have excess flux in the UV, connected to its binarity. They are confirmed spectroscopic binaries, and their hot companions are compact objects, likely to be sdA stars. Based on these results, we conclude that they are products of the binary mass transfer.
    \item From comparing the position of the CSPN with the theoretical pAGB evolutionary tracks, we found that it has entered the WD cooling phase, and its mass is found to be $\sim0.66M_{\odot}$. The mass of the progenitor corresponding to the WD of mass $0.66M_{\odot}$ would be $\sim2.1M_{\odot}$, similar to the turn-off mass of the cluster, further confirming its membership.
\end{itemize}

\section*{Acknowledgements}
We thank the anonymous referee for the valuable comments and suggestions. AS acknowledges support from SERB Power Fellowship. S. Rani wants to thank Vikrant Jadhav for providing the field ELM WDs SED fit parameters catalog. S. Rani thanks Sonith L. S. for the fruitful discussions. This publication utilizes the data from {\it AstroSat} mission's UVIT, which is archived at the Indian Space Science Data Centre (ISSDC). The UVIT project is a result of collaboration between IIA, Bengaluru, IUCAA, Pune, TIFR, Mumbai, several centers of ISRO, and CSA. This research made use of VOSA, developed under the Spanish Virtual Observatory project supported by the Spanish MINECO through grant AyA2017-84089. This research also made use of the Aladin sky atlas developed at CDS, Strasbourg Observatory, France \citep{Bonnarel2000}.

\vspace{5mm}


\software{GaiaTools \citep{2019MNRAS.484.2832V}, Topcat \citep{2011ascl.soft01010T}, Matplotlib \citep{2007CSE.....9...90H}, NumPy (\citealp{2011CSE....13b..22V}), Scipy \citep{2007CSE.....9c..10O, article}, Astropy \citep{2013A&A...558A..33A, 2018AJ....156..123A} and Pandas \citep{mckinney-proc-scipy-2010}}



\bibliographystyle{aasjournal}





\end{document}